# Asymmetrical Defect Sink Behaviour of HCP/BCC Zr/Nb Multilayer Interfaces: Bubble-Denuded Zones at Nb Layers

N. Daghbouj [a*], H.S. Sen [b*], M. BenSalem [a], J.Duchoň[c], B. Li [d], M. Karlík [e], F. Ge [f], V. Krsjak [g], P. Bábor[h], M.O. Liedke[i], M. Butterling[j], A. Wagner [i], B. Karasulu[b], T. Polcar[ak]

[a]Department of Control Engineering, Faculty of Electrical Engineering, Czech Technical University in Prague, Technická 2, 160 00 Prague 6, Czechia

[b]Department of Chemistry, University of Warwick, Coventry CV4 7AL, U.K.

[c]Institute of Physics of the Czech Academy of Sciences, Na Slovance 1999/2, 182 21 Prague 8, Czechia

[d]State Key Laboratory for Environment-friendly Energy Materials, Southwest University of Science and Technology, Mianyang, Sichuan 621010, China

[e]Department of Materials, Faculty of Nuclear Sciences and Physical Engineering, Czech Technical University in Prague, Trojanova 13, 120 00 Prague 2, Czechia

[f]Laboratory of Advanced Nano Materials and Devices, Ningbo Institute of Materials Technology and Engineering, Chinese Academy of Sciences, Ningbo 315201, China

[g]Institute of Nuclear and Physical Engineering, Faculty of Electrical Engineering and Information Technology, Slovak University of Technology, Ilkovicova 3, 812 19 Bratislava, Slovakia

[h]CEITEC - Central European Institute of Technology, Brno University of Technology, 616 00 Brno, Czech Republic

[i]Institute of Radiation Physics, Helmholtz-Zentrum Dresden-Rossendorf, Bautzner Landstr. 400, 01328 Dresden, Germany

[j]Reactor Institute Delft, Department of Radiation Science and Technology, Faculty of Applied Sciences, Delft University of Technology, Mekelweg 15, NL-2629 JB Delft, The Netherlands

[k]School of Engineering, University of Southampton, Southampton SO17 1BJ, United Kingdom

**Abstract**

Radiation-induced helium (He) bubble formation poses a major challenge to the structural integrity of materials in nuclear energy systems. In this study, we investigate defect evolution and He behavior in Zr/Nb nanoscale metallic multilayers (NMMs) with immiscible BCC/HCP interfaces, irradiated with 80 keV He ions at fluences ranging from $1\times10^{16}$ to $1\times10^{17}$ He/cm². For comparison, single-crystal Nb and polycrystalline Zr were also irradiated under identical conditions to serve as reference materials. Using cross-sectional TEM, SIMS, STEM-EELS, nanoindentation, Doppler Broadening Positron Annihilation Spectroscopy (DBPAS), Positron Annihilation Lifetime Spectroscopy (PALS), and atomistic simulations (DFT and MD), we reveal a highly asymmetric damage response across the multilayer interfaces. Zr layers exhibit larger He bubbles (1.5–2.8 nm), higher swelling (~1.2%), and greater helium retention, while Nb layers develop bubble-denuded zones (BDZs) exclusively around the interfaces, where bubble nucleation is strongly suppressed and swelling is limited to ~0.4%. This asymmetry arises from differences in atomic transport properties: DFT calculations show lower migration barriers for vacancies and He atoms in Nb (0.4 and 0.19 eV, respectively), enabling efficient defect migration and recombination at interfaces, whereas Zr retains defects due to higher migration barriers. EELS and DBS-PALS measurements confirm bubble densities of 63–96 He/nm³ and the presence of sub-nanometer open volumes. Compared to monolithic samples, the Zr/Nb multilayers exhibit ~50% lower irradiation-induced hardening and reduced He retention (11% vs. 17.5% in single-crystal Nb and 16% in polycrystalline Zr). These findings highlight the role of interfaces in driving asymmetric radiation damage and demonstrate the effectiveness of BCC Nb layers in mitigating defect growth. Overall, Zr/Nb multilayers are established as a superior alternative to conventional single- and polycrystalline materials for extreme irradiation environments.

* Corresponding authors' e-mails: daghbnab@fel.cvut.cz, huseyin-sener.sen@warwick.ac.uk

## 1. Introduction

Nuclear energy remains a cornerstone of modern low-carbon energy strategies. However, realizing its full potential requires the development of advanced structural materials that can withstand the extreme conditions of nuclear reactors, including high temperatures, radiation, and prolonged neutron exposure [1-3]. A critical issue is the accumulation of helium (He) atoms, which results from transmutation reactions induced by neutron irradiation [2]. In fusion reactors, alongside displacement damage, high concentrations of He atoms are generated through (n, α) and other transmutation reactions. Due to their low solubility in most metals, He atoms readily aggregate to form nanoscale bubbles within irradiated structural materials [4]. These bubbles, often nucleating at defect sites such as vacancies, evolve into larger voids, contributing to swelling, hardening, and embrittlement — all of which degrade mechanical and thermal performance [5–11]. These issues severely compromise the mechanical and thermal properties of nuclear materials, ultimately limiting their operational lifespan and safety [12, 13]. The damage caused by irradiation is primarily driven by the aggregation and evolution of high-density irradiation defects. The resulting ultrahigh-irradiation hardening observed in these materials restricts their applicability in advanced nuclear reactors [2, 14]. It is well-documented that He atoms produced by nuclear transmutation reactions are poorly soluble in metals and tend to be trapped by vacancies, forming He bubbles [15-17]. Over time, these pressurized bubbles can grow, creating voids that further embrittle the irradiated materials [18-20]. Addressing the challenges posed by irradiation damage is critical for improving nuclear materials' durability and safety in current and future reactor designs. One effective strategy to mitigate these issues involves increasing material defect sinks, such as grain boundaries (GBs), twin boundaries (TBs), free surfaces, and heterogeneous interfaces [21-26]. These sinks trap irradiation-induced defects, promoting their annihilation and controlling their distribution, thereby mitigating irradiation damage [26-30]. To mitigate these effects, researchers have explored strategies such as grain refinement [31-35] and, more recently, the design of nanostructured materials with high-density heterogeneous interfaces [36, 37]. These architectures enhance defect recombination and He trapping, offering a promising path to radiation-tolerant materials. As a result, the design of nanostructures, particularly nanostructured metallic multilayers (NMMs) with a high density of heterointerfaces, has emerged as a promising approach to enhance radiation tolerance [24]. Heterostructured NMMs, including FCC/FCC systems like Ag/Ni [38] and Cu/Ag [39], FCC/BCC systems such as Ag/V [40], Cu/Ta [41], Cu/Nb [42-45] and Cu/Mo [46], FCC/HCP systems like Cu/Co [47] and Cu/Zr [48], BCC/BCC systems such as Fe/W [49], BCC/HCP systems like Mo/Zr [50] and Nb/Zr [51-54], and even crystalline/amorphous systems [55-58], have garnered significant attention for their unique mechanical properties and irradiation damage behavior. Studies indicate that systems with more interfaces (or smaller layer thickness) exhibit less irradiation hardening [47]. However, the sink efficiency of different interfaces varies widely, as evidenced by numerous studies [44-47]. Understanding and optimizing these interfacial behaviors are essential for advancing the development of radiation-tolerant materials for nuclear applications.

The design of multilayers with high-density heterophase interfaces to mitigate He irradiation damage is grounded in the principle that interfaces can efficiently trap He atoms and vacancies while providing abundant nucleation sites for He bubbles [6]. Research on He-irradiated multilayers has consistently demonstrated the critical role of heterophase interfaces in enhancing radiation tolerance. For instance, Li et al. [42] investigated He bubble size and distribution in Cu and 5 nm Cu/Nb multilayers after 7 at.% He irradiation. Their findings revealed that the average bubble radius in Cu/Nb multilayers was approximately 0.5 nm, significantly smaller than the 1.35 nm observed in pure Cu. Furthermore, the bubble volume fraction in Cu/Nb multilayers was only about 1.74%, compared to 5.77% in Cu. Similar results were observed in 6 nm V/Ag multilayers, where He bubbles were smaller in size and lower in volume fraction compared to pure Ag after He irradiation [59].

Additional studies have demonstrated that defect density generally decreases with reduced layer thickness [50, 60-63], although some thin multilayers exhibit larger-sized defects, such as bubbles [64-66]. Interestingly, in certain systems, such as He-irradiated Cu/W, denser He bubbles are observed in Cu compared to W [67]. The presence of bubble-denuded zones (BDZs) in multilayers further highlights the critical role of heterophase interfaces in suppressing irradiation-induced defects [21]. While BDZs have been extensively studied in materials with grain boundaries (GBs), their behavior in multilayer systems remains less explored. In Cu, for example, studies suggest that GBs can absorb nearby He atoms and vacancies, creating regions with fewer He bubbles on either side of the GBs [68, 69]. However, the mechanisms underlying the formation of BDZs and the variations in bubble size and density across different multilayer systems are still poorly understood and require further investigation. A deeper understanding of these phenomena is essential for advancing the design of radiation-tolerant materials, particularly for applications in advanced nuclear reactors.

Extensive research has been conducted on the radiation behavior of multilayers, especially FCC/FCC and FCC/BCC systems, providing important insights into their microstructure-property relationships under irradiation [51, 70-80]. More recently, BCC/HCP multilayers, such as Nb/Zr [76, 77], have garnered increasing attention for their potential in nuclear applications. While many studies have investigated irradiated multilayers, there is still a lack of detailed understanding of helium-induced damage evolution and asymmetric defect responses in immiscible BCC/HCP systems, particularly under high He flux. Additionally, comparative assessments with monolithic references are limited, making it challenging to isolate and understand the interfacial effects. These gaps motivate the present study, which focuses on helium-induced damage in Zr/Nb multilayers and benchmarks the behavior against single-crystal Nb and polycrystalline Zr. The Zr-Nb alloy is a highly promising material for nuclear applications due to its excellent corrosion resistance, high-temperature strength [81, 82], and Despite extensive studies on He bubble behavior in metallic NMMs, the immiscible BCC/HCP Nb/Zr system (Hmix = +4 kJ mol$^{-1}$) was chosen to investigate the role of interfaces in the behavior of He and point defects. Specifically, this research aims to enhance the understanding of BCC/HCP interfaces in radiation damage evolution under high He flux. Despite extensive studies on He bubble behavior in metallic multilayers, the directional role of interface asymmetry, particularly in immiscible systems like Zr/Nb, remains unclear. Prior work has demonstrated the formation of bubble-denuded zones (BDZs) in Cu/Nb and Fe/W systems, but the impact of crystallographic contrast (BCC/HCP), defect mobility, and helium transport across such interfaces has not been fully explored. Moreover, comparative studies with single-crystal and polycrystalline reference materials are lacking, limiting the practical relevance of current models. In this study, we systematically investigate helium-induced damage evolution in Zr/Nb nanoscale multilayers irradiated with 80 keV He ions, and benchmark the results against monolithic single-crystal Nb and polycrystalline Zr. Using advanced microscopy, positron spectroscopy, and atomistic simulations, we reveal an asymmetric defect response across the interface and identify exclusive formation of BDZs at the Nb side. Our findings offer mechanistic insight into interfacial defect transport and establish a framework for designing radiation-tolerant materials through interface engineering.

## 2. Materials and methods

2.1. Experimental setup

2.1.1. Sample Structure

High-purity zirconium (Zr) and niobium (Nb) targets were used to deposit Zr/Nb nanomultilayers (NMMs) with a bilayer period of 96 nm (57 nm Zr + 39 nm Nb) onto single-crystal (111) silicon (Si) substrates by DC magnetron sputtering. The total film thickness ranged from 1.5 to 2 μm. Detailed fabrication procedures are reported elsewhere [52, 80]. The deposited multilayers were sectioned into smaller pieces for ion implantation at varying fluences. For comparison, single-crystal (110) Nb and polycrystalline Zr samples were also prepared to evaluate the effect of interfaces. The chosen multilayer architecture ensured sufficiently thick Zr and Nb layers to allow distinct helium bubble nucleation and growth within each layer, while also facilitating detailed cross-sectional TEM and STEM characterization of bubble distributions relative to the interfaces. Furthermore, the design parameters match those of previous studies [52, 80], enabling direct comparison. The total film thickness (1.5–2 μm) guarantees that multiple bilayers lie within the helium implantation depth. Bulk Nb reference samples consisted of single-crystal (110) substrates purchased from Goodfellow, while bulk Zr references were polycrystalline with grain sizes exceeding 300 nm, as confirmed by TEM and XRD (Fig. S1). These Zr references originated from UJP Praha, a.s., Czech Republic. The multilayer films exhibited grain sizes larger than the individual layer thicknesses and strong crystallographic textures of (0002) Zr and (110) Nb, as revealed by automated crystal orientation mapping (ACOM) and XRD analysis (Fig. S1).

2.1.2. Ion Irradiation

Helium ion implantation was carried out at room temperature with an ion energy of 80 keV at three different fluences: low fluence (LF = 1 x $10^{16}$ He/cm$^2$), medium fluence (MF = 5 x $10^{16}$ He/cm$^2$), and high fluence (HF = 1 x $10^{17}$ He/cm$^2$). According to SRIM (Stopping and Range of Ions in Matter) simulations [83], for the ZrNb$_{96}$ multilayers, the peak He concentrations were 1 at.% (LF), 5 at.% (MF), and 12 at.% (HF), corresponding to radiation damage levels of 0.3, 1.4, and 2.9 displacements per atom (dpa), respectively, at a projected range (Rp) of 355 nm (see Fig. 1). SRIM results indicate that the peak He concentration occurs at a greater depth than the peak radiation damage. The He penetration depths in pure elements vary: 250 nm in Nb and 380 nm in Zr. The damage profiles were calculated using the "Quick" Kinchin-Pease model, following the guidelines of Stoller et al. [84], with displacement threshold energies (Ed) set at 40 eV for Zr and 60 eV for Nb.

2.1.3. Secondary Ion Mass Spectrometry (SIMS)

SIMS depth profiling was conducted using a Cameca IMS 7f instrument to detect positively charged secondary ions. A 1 keV $Cs^+$ primary ion beam at an incidence angle of 45° and a current of ~105 nA was used. Secondary $CsHe^+$ ions were collected to overcome He's high first ionization potential. The raster area was 100 μm × 100 μm, with a 33 μm-diameter analysis region centered within the crater. Depth calibration was achieved using stylus profilometry. The sputtering rate was normalized to the average primary ion current, and the He concentration scale was calibrated using the known LF He fluence into Nb. Each sample was measured twice, and secondary ions ($^4He^+$, $^{90}Zr^+$, and $^{93}Nb^+$) were recorded. SIMS profiles obtained at HF for Zr/Nb, Zr, and Nb are shown in Fig. 1; LF profiles are in Fig. S2. These data, along with SRIM simulations, reveal contrasting damage distributions and He penetration depths between Zr and Nb, offering insights into defect evolution.

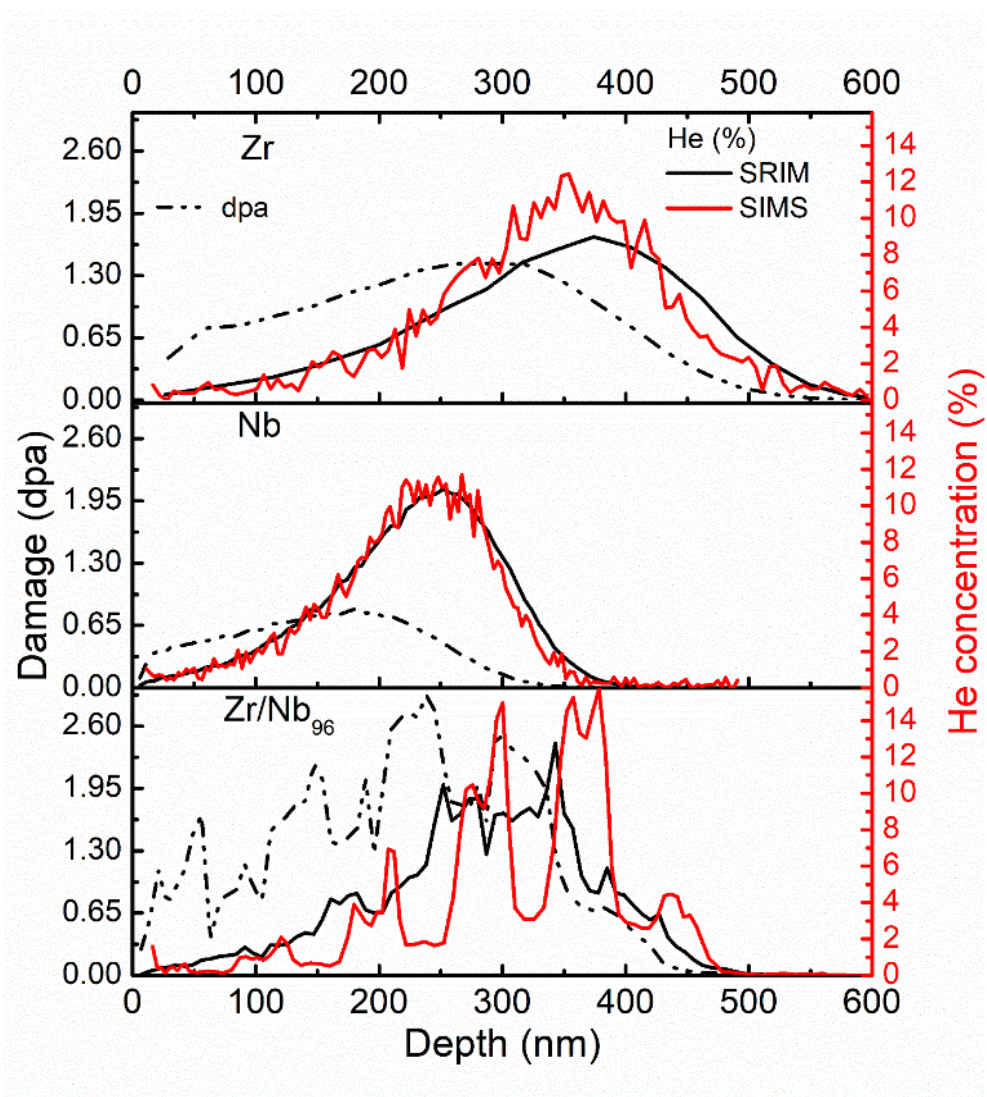

Fig. 1. Ion concentration profiles (in at.%) predicted by SRIM and measured by SIMS (right axis), along with the corresponding damage profile (dashed line, left axis) in (a) Zr, (b) Nb, and (c) the fabricated Zr/Nb multilayer irradiated with 80 keV He ions at a fluence of $1 \times 10^{17}$ He$^+$/cm² (HF).

2.1.4. Transmission Electron Microscopy (TEM)

Cross-sectional scanning transmission electron microscopy (STEM) and energy-dispersive X-ray spectroscopy (EDX) were performed using a JEM-2200 FS microscope operated at 200 kV to characterize elemental distribution and interface quality. TEM lamellae were prepared via focused ion beam (FIB) milling in a Helios 600 dual-beam system. Initial trenching and thinning were done with a 30 kV Ga$^+$ beam, followed by final thinning with a 2 kV beam. A Pt layer was deposited for surface protection. Sample thickness was measured by electron energy loss spectroscopy (EELS) and found to be ~90 nm. Orientation analysis was conducted using automated crystal orientation mapping (ACOM) in a FEI Tecnai TF20 TEM equipped with a NanoMegas ASTAR precession unit [8]. The ACOM used a 7 nm probe, 0.7° precession angle, and 65 mm camera length. STEM-EELS spectra were acquired using a GATAN GIF 2001 system, with ~0.7 eV resolution, convergence/collection angles of 5.86 mrad and 14.68 mrad, a beam current of 9.87 nA, and a probe size around 0.35 nm. Acquisition was optimized to enhance ³He K-edge signal, while avoiding drift and He detrapping. HAADF-STEM images were taken concurrently. Bubble size distributions were analyzed using a developed algorithm in MATLAB® as shown in Fig. S3.

2.1.5. VariableEnergy Positron Annihilation Spectroscopy (VE-PAS)

Variable-energy PALS was conducted at the ELBE facility (Helmholtz-Zentrum Dresden-Rossendorf, Germany) using the Monoenergetic Positron Spectroscopy (MePS) beamline, with a positron flux of ~$10^6$/s [85]. Positron implantation energies ranged from 2 to 12 keV, and lifetime spectra were collected at each energy level. Each spectrum included ≥$10^7$ counts to ensure statistical reliability. Deconvolution was performed using the PALSfit software package [86], which applied a non-linear least-squares fitting method, minimized by the Levenberg-Marquardt algorithm, revealing 2-3 lifetime components. Positrons, due to their positive charge, are repelled by atomic nuclei and become trapped in open-volume defects (vacancies, voids), where they exhibit longer lifetimes due to reduced electron density [87,88]. Upon annihilation, two 511 keV γ-rays are emitted, and the time difference between injection and annihilation yields the positron lifetime spectrum. This technique enables detection of defect types and concentrations [89,90]. Doppler broadening variable energy positron annihilation spectroscopy (DB-VEPAS) measurements have been conducted at the slow positron beamline (SPONSOR) [91]. Positrons have been implanted into samples in the range 0.05-35 keV, which realizes depth profiling. A mean positron implantation depth $\langle z \rangle$ was approximated using a material density ($\rho$) dependent formula [92]:

$$\langle z \rangle \text{ [nm]} = \frac{36}{\rho \text{ [g·cm}^{-3}\text{]}} E_p^{1.62} \text{ [keV]}$$

At the annihilation site, thermalized positrons have very small momentum compared to the electrons, hence a *broadening of the 511 keV line* is observed mostly due to the momentum of the electrons. The width variations are measured with a high-purity Ge detector (energy resolution of 1.1 ± 0.1 keV). The broadening is described by two distinct parameters, *S* and *W*, defined as a fraction of the annihilation line in the middle and outer regions, respectively and normalized to the total area below the curve. The S-parameter is a fraction of positrons annihilating with low-momentum valence electrons and represents vacancy-type defects and their concentration. The W-parameter approximates the overlap of the positron wavefunction with high-momentum core electrons and is a fingerprint of the local atomic chemistry [93].

2.1.6. Nanoindentation

Nanoindentation experiments were performed using a Hysitron TI 950 TriboIndenter equipped with a three-sided Berkovich diamond tip. Continuous stiffness measurement (CSM) mode was employed to continuously record hardness and modulus as a function of indentation depth. For each sample, at least 16 indents were performed to ensure statistical reliability of the measured values. When the indentation depth range between 300 and 500 nm, Pile-up and sink-in effects were examined using post-indentation atomic force microscopy (AFM). This analysis confirmed negligible pile-up or sink-in around indents within the selected depth range (Fig. S4). Surface roughness was also quantified by AFM before and after irradiation. The measured root-mean-square (RMS) roughness values ranged from 6 to 8 nm for Zr/Nb multilayers and 9 to 14 nm for pure Zr and Nb samples. Since these roughness levels are at least an order of magnitude smaller than the indentation depths, they did not significantly influence the hardness data. This conclusion is supported by the fact that the plastic deformation zone generated during indentation extends well beyond the surface asperities, ensuring the reliability of the mechanical property measurements.

## 2.2. Atomistic Simulations

This study employed two atomistic simulation techniques: classical molecular dynamics (MD) and Density Functional Theory (DFT). To explore the irradiation response of Nb/Zr materials, MD simulations were performed using the primary knock-on atom (PKA) method [94] within the LAMMPS [95] framework. The interatomic interactions in Zr-Nb metallic multilayers were described by the Angular Dependent Potential (ADP) developed by Smirnova and Starikov [96], which is parameterized based on ab-initio data. The potential was used in our previous work [79]. However, this potential does not include He atoms. It accurately replicates point defect formation and diffusion in both HCP Zr and BCC Nb crystal structures. The study investigated bulk and interface regions separately to determine the interface's role in damage evolution. Bulk simulations used supercells of Zr and Nb, each containing approximately 50,000 atoms, whereas interface simulations involved a (110)Nb/(0002)Zr interface system measuring 10 nm x 10 nm x 40 nm, with each layer having a thickness of about 20 nm. A brief description of the model used in interface simulations can be found in S5. The simulations covered temperatures from 300 K to 750 K. Once equilibrium had been reached, a primary knock of 3 MeV kinetic energy was applied to a selected atom to replicate an alpha particle collision. The velocity direction was set perpendicular to the interface, either towards or away from it, ensuring a direct collision with the nearest neighboring atom. This approach prevents excessive atomic displacement before impact, allowing for a focused study of radiation damage near the interface. Given the high initial velocities, a variable time-step strategy was implemented to avoid unphysical atomic jumps: 1 as for the first 50,000 steps, 10 as for the following 45,000 steps, 100 as for the next 95,000 steps, and 1 fs for the final 50,000 steps, resulting in a total simulation time of 60 ps. OVITO software [97] was used for defect identification and analysis.

For simulations incorporating He atoms within the supercell, plane-wave DFT calculations were conducted using the Vienna Ab Initio Simulation Package (VASP, v. 6.3) [98], which applies periodic boundary conditions and the pseudopotential approximation. The calculations utilized the projector-augmented wave (PAW) method [99] combined with the Perdew-Burke-Ernzerhof (PBE) version of the Generalized Gradient Approximation (GGA) for exchange–correlation potentials [100]. Atomic positions and lattice parameters were fully relaxed using the conjugate-gradient method until residual forces are below 0.01 eV/Å. The Brillouin zone was sampled using a Γ-centered Monkhorst–Pack (MP) grid with a k-point spacing finer than 0.1 Å$^{-1}$. A plane-wave energy cut-off of 520 eV was adopted for both cell and geometry optimizations as well as NEB [101] energy barrier calculations, while a reduced cut-off of 400 eV was used for Ab Initio Molecular Dynamics (AIMD) simulations. Both types of simulations employed the "normal" precision setting (PREC=Normal), an electronic convergence threshold of $10^{-8}$ eV, and a Gaussian smearing width of 0.1 eV. No symmetry constraints were applied during the dynamic simulations.

## 3. Results

### 3.1. Microstructural evolution of irradiated Zr, Nb, and Zr/Nb$_{96}$ multilayers

Following high-fluence He ion irradiation, helium bubbles form in both polycrystalline Zr (Fig. 2a,b) and single-crystalline (110) niobium (Nb) (Fig. 3a,b), with distinct differences in bubble morphology. In Zr, larger bubbles and higher swelling are observed, peaking at ~300 nm depth, while Nb exhibits smaller bubbles and reduced swelling, peaking at ~250 nm. Notably, Zr/Nb$_{96}$ nanomultilayers (NMMs) reveal asymmetric behavior (Fig. 4a-c): denser and larger bubbles form in Zr layers, whereas bubble-denuded zones (BDZs) appear near Nb interfaces, suppressing bubble nucleation (highlighted by dashed line in Fig. 4 b). This contrast underscores the role of heterophase interfaces in mitigating radiation damage. XTEM analysis, conducted under ~400 nm under-focus conditions to enhance contrast, quantified bubble characteristics by averaging measurements across five regions at each depth. Bubble diameter, surface coverage, density, and swelling were calculated using bright-field TEM images. Swelling (*S*) was determined via the equation (1).

$$S = \frac{V_{bub}}{V_0 - V_{bub}} \times 100\ \% = \frac{\sum_{i=1}^{N} \frac{4\pi r_i^3}{3}}{A \times \delta - \sum_{i=1}^{N} \frac{4\pi r_i^3}{3}} \times 100\ \% \qquad (1)$$

where $V_{bub}$ is the total bubble volume, and $V_0$ is the unirradiated material volume, A is the TEM image area of layer *i* and $\delta$ is the layer thickness. Additional quantitative details and spatial distributions of bubbles are presented in Figs. S6-S8.

In all cases, bubble diameter, surface coverage, and density peak at depths where He concentration maximizes (~300 nm for Zr, ~250 nm for Nb, and ~350 nm for Zr/Nb$_{96}$), aligning with SIMS profiles (Figs. 2–4c,d). However, Zr/Nb$_{96}$ multilayers exhibit a threefold lower peak bubble density than Nb and twofold lower than Zr, alongside smaller bubbles in Nb layers compared to single-crystalline Nb. Critically, swelling in Nb layers decreases by three times compared to pure Nb, directly implicating Zr/Nb interfaces in defect recombination. This interfacial effect highlights the potential of engineered multilayers to enhance radiation tolerance, particularly in Nb, by leveraging interfaces as defect sinks.

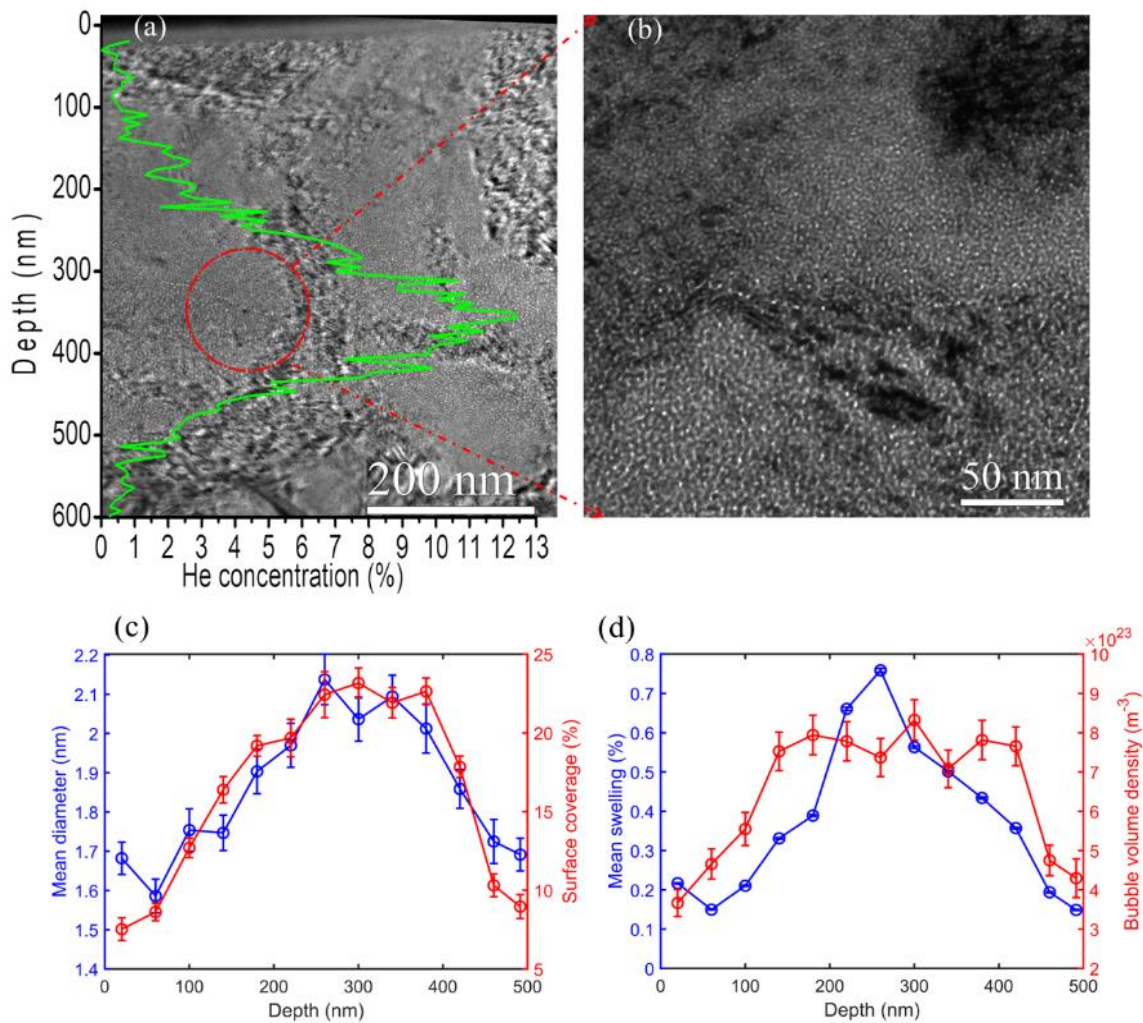

Fig. 2. (a) XTEM image of a polycrystalline Zr sample after He ion irradiation at high fluence (HF = $1\times10^{17}$ He/cm²), overlaid with the He concentration profile measured by SIMS (green line). (b) High-magnification view of the region corresponding to the peak He concentration at a depth of 350 nm below the sample surface. (c) Mean diameter of He bubbles and the surface area occupied by He bubbles as a function of depth along the sample's normal direction. (d) Mean swelling and bubble volume density as a function of depth along the normal direction beneath the surface.

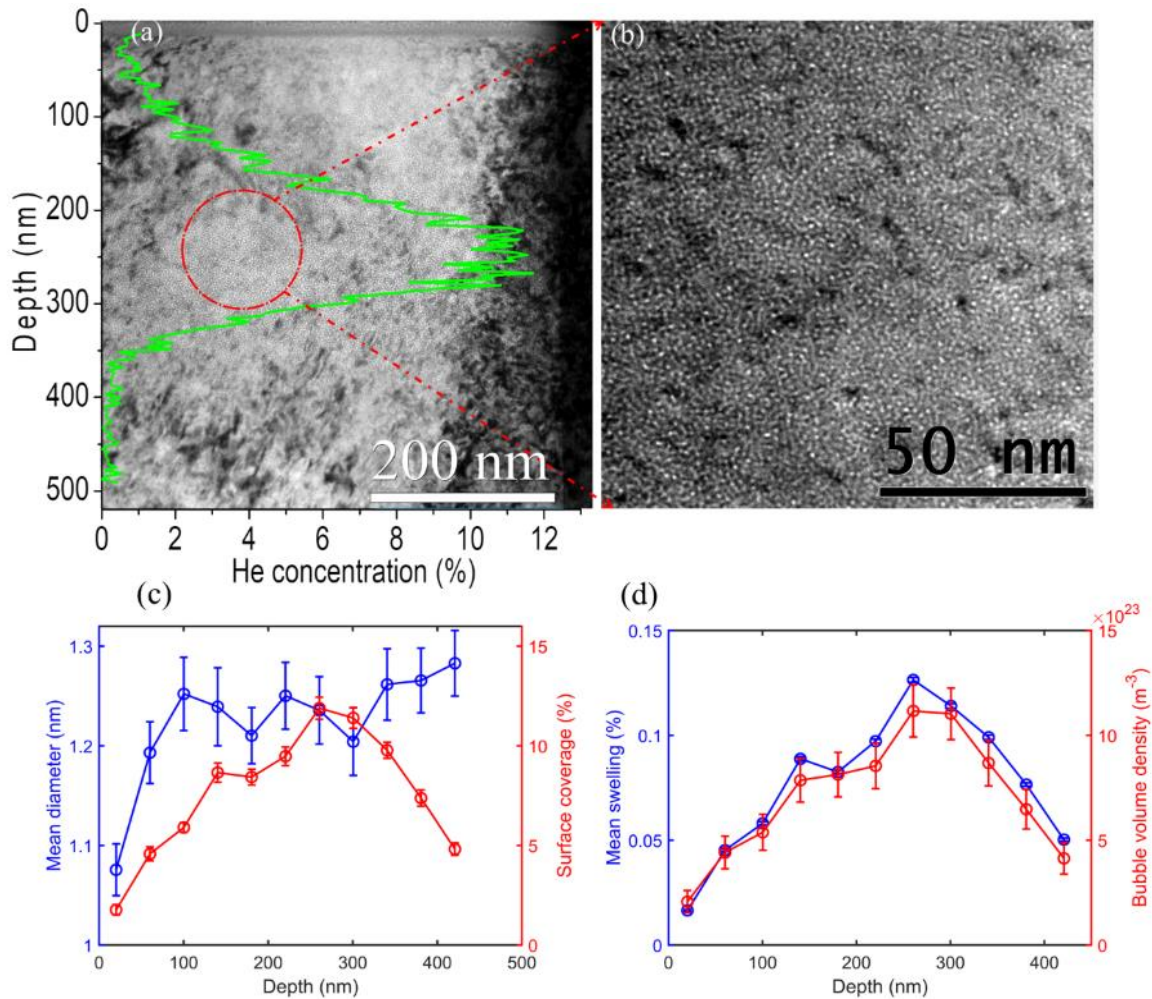

Fig. 3. (a) XTEM image of a single crystalline (110)Nb sample after He ion irradiation at high fluence (HF = $1\times10^{17}$ He/cm²), overlaid with the He concentration profile measured by SIMS (green line). (b) High-magnification view of the region corresponding to the peak He concentration at a depth of 250 nm below the sample surface. (c) Mean diameter of He bubbles and the surface area occupied by He bubbles as a function of depth along the sample's normal direction. (d) Mean swelling and bubble volume density as a function of depth along the normal direction beneath the surface.

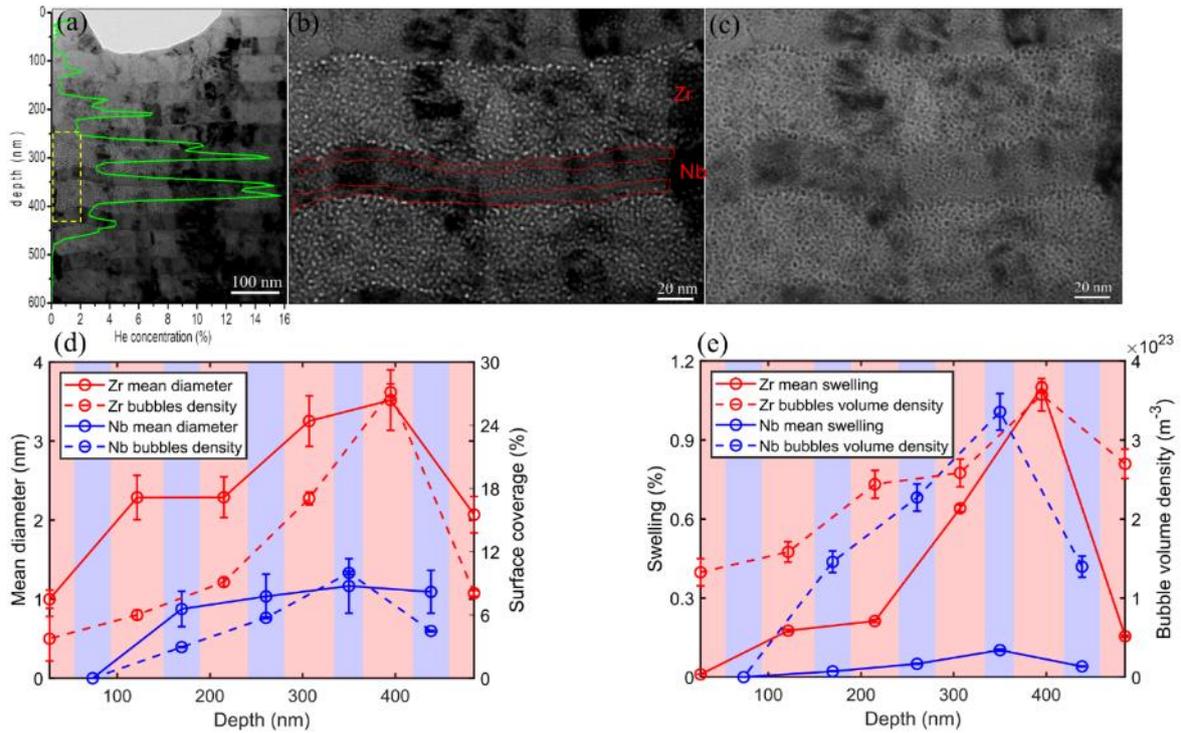

Fig. 4. (a) XTEM image of He ion irradiated Zr/Nb$_{96}$ NMMs at high fluence (HF = 1×10$^{17}$ He/cm²), overlaid with the He concentration profile measured by SIMS (green line). (b, and c) High-magnification view of the region corresponding to the peak He concentration at a depth of 350 nm below the sample surface in over- and under-focus, respectively. (d) Mean diameter of He bubbles and the surface area occupied by He bubbles in Zr and Nb layers as a function of depth along the sample's normal direction. (e) Mean swelling and bubble volume density in Zr and Nb layers as a function of depth along the normal direction beneath the surface.

3.2. Irradiation Hardening After He Irradiation at 1 at%, 5 at%, and 12 at%

Irradiation-induced defects impede dislocation motion, resulting in material hardening, a critical factor in evaluating irradiation resistance [70]. Nanoindentation was employed to assess the depth-dependent hardness profiles of Zr polycrystals, (110) Nb single crystals, and Zr/Nb$_{96}$ nanoscale metallic multilayers (NMMs) before and after helium irradiation at low (LF: 1 at%), medium (MF: 5 at%), and high (HF: 12 at%) concentrations, as illustrated in Fig. 5(a–c). For the Zr/Nb$_{96}$ multilayers, hardness was determined at a penetration depth corresponding to approximately 10% of the $_{total}$ film thickness, indicated by the vertical dashed line in Fig. 5c. In contrast, the intrinsic hardness of Zr and Nb samples was calculated using Nix and Gao's model [102], following the methodology proposed by Kasada et al.

[103]. This model accounts for the indentation size effect and relates nanohardness ($H$) to indentation depth ($h$) through the equation (2).

$$H = H_0 \sqrt{1 + \frac{h^*}{h}} \tag{2}$$

Where $H$ is the hardness at the indentation depth of h, $h^*$ denotes the characteristic length depending on the shape of the indenter and material type, and $H_0$ represents the real hardness at infinite depth.

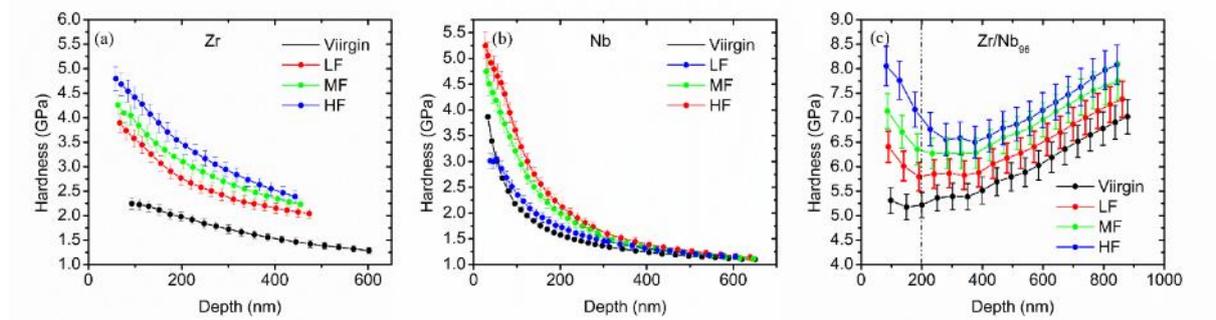

Fig. 5. (a-c) Average hardness (H) versus the indentation depth (h) of (a) Zr ploycrystal, (b) (110)Nb, and (c) Zr/Nb$_{96}$ NMMs before and after He ion irradiation

According to Eq. (1), the raw data can be plotted as the square of hardness ($H^2$) and the reciprocal of the indentation depth (1/h), as shown in Fig. 6(a-d), and the intercepts after curve fitting are the square of the real hardness ($H_0$).

Fig.6. The fitting results using the Nix-Gao model $H^2$ vs $\frac{1}{h}$ of Zr polycrystal before and after He irradiation; (a-d) correspond to virgin, LF, MF, and HF, respectively. The fitting of nanoindentation data was conducted at indentation depths ranging from 70 nm to 120 nm, corresponding to a regime where the plastic zone fully encompasses the irradiated region. Analogous fittings for Nb and multilayers are provided in the Supplementary Information. Deviations observed at shallow depths are attributed to surface roughness and surface deformation effects that deviate from ideal model assumptions. This ensures that the measured mechanical response reflects the entire damage layer. The fitting results for both virgin and irradiated single-crystal Nb are presented in Fig. S9. For the Zr/Nb$_{96}$ multilayers, hardness values for both the unirradiated and irradiated states were extracted at an indentation

depth approximately ten times the total multilayer film thickness. This depth criterion minimizes substrate effects and ensures that the measurement represents the effective mechanical properties of the entire multilayer structure, rather than those of individual layers or the underlying substrate.

The irradiation hardening is calculated using equation (3).

$$\Delta H = H_{irrad} - H_{virgin} \tag{3}$$

where $H_{irrad}$ and $H_{virgin}$ represent the hardness of irradiated and unirradiated materials, respectively. The results are summarized in Fig. 7.

As the helium dose increases from 1 at.% to 12 at.%, irradiation hardening rises in all three samples. However, the Zr/Nb$_{96}$ multilayers exhibit significantly lower hardening compared to pure Zr and Nb. At a helium peak concentration of ~7 at.%, the hardening magnitude in Zr/Nb$_{96}$ is approximately half that observed in Cu/Nb$_{70}$ [42] and also lower than in Fe/W$_{50}$ systems [49]. This reduced hardening highlights the role of hcp/bcc interfaces in suppressing defect accumulation and dislocation pinning, thereby enhancing irradiation resistance.

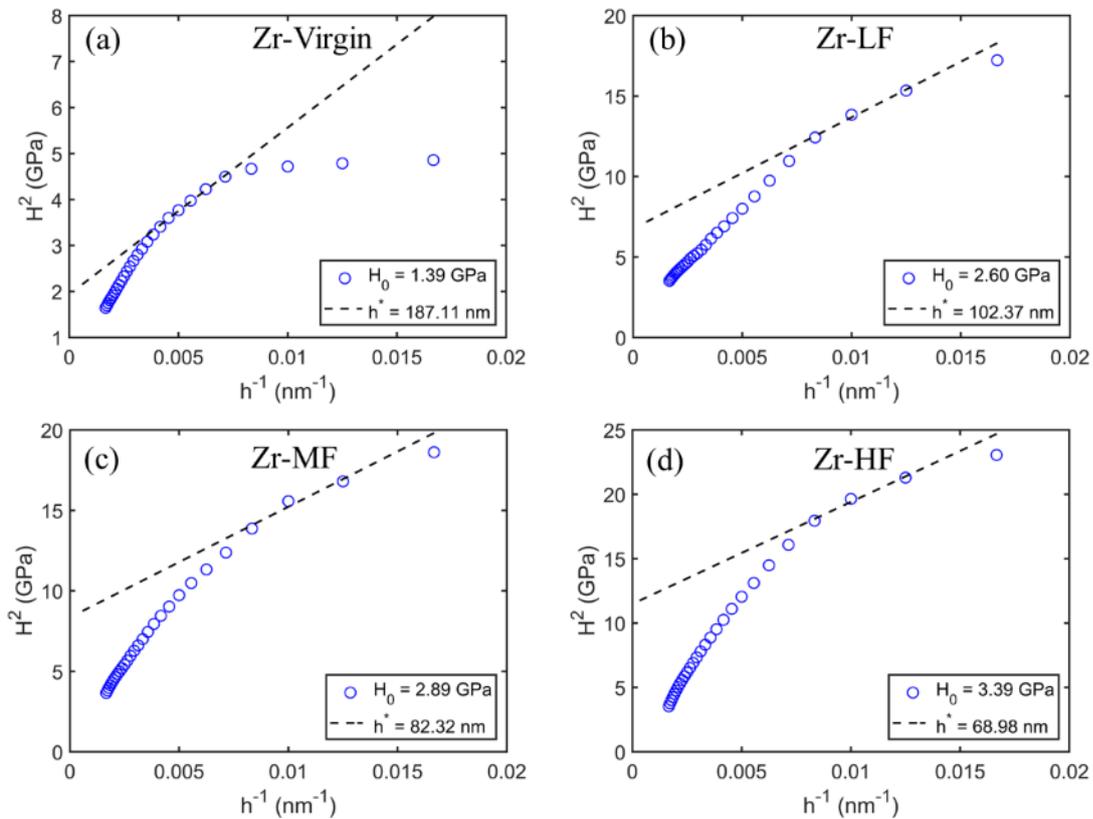

Fig.6. He fluence-dependent radiation-hardening in pure Zr, Nb, and Zr/Nb$_{96}$ multilayers after He irradiation with low (LF), medium (MF), and high fluence (HF).

TEM analysis reveals a striking contrast in radiation response: HCP Zr exhibits pronounced susceptibility to helium bubble growth, while BCC Nb demonstrates superior defect tolerance. Bubble-denuded zones (BDZs) near Nb interfaces significantly suppress swelling, whereas Zr interfaces remain active sites for bubble nucleation. This asymmetric behavior in Zr/Nb$_{96}$ multilayers highlights the critical role of interfacial engineering in nuclear materials design. By bridging microstructural observations, such as bubble size and density, with mechanistic insights into interface-mediated defect absorption, this study establishes a transformative framework for optimizing radiation-resistant materials. To unravel the atomic-scale origins of this behavior, DFT and cascade simulations are conducted, as detailed in Section 4.2. These computational efforts complement experimental findings, offering a holistic strategy to tailor interfaces for extreme nuclear environments.

## 4. DISCUSSION

4.1 Estimation of Helium Concentration Contributing to Bubble Formation

In irradiated materials, the gas pressure within a bubble often counterbalances the osmotic pressure at the surface. Under thermodynamic equilibrium conditions, this pressure $p$ is related to the surface energy $\gamma$ and the bubble radius $r$ by the equation (4).

$$p = 2\gamma/r \qquad (4)$$

This relationship assumes a stable equilibrium between the bubble and the surrounding material [63]. The surface energies for pure Zr and Nb are taken as samples, where $\gamma_{Zr} = 1.73 \, J \, m^{-2}$, $\gamma_{Nb} = 2.36 \, J \, m^{-2}$ [104]. Using bubble size statistics from TEM images (Figs 2, 3, and 4), the average bubble pressure $\bar{p}$ as a function of depth was calculated as shown in equation (5).

$$\bar{p} = \frac{\sum_i^N p_i}{N} \qquad (5)$$

Pure Zr: 4 to 5.5 GPa, Pure Nb: 7.5 to 9.5 GPa, Zr layers in Zr/Nb NMMs: 2 to 3.5 GPa, and Nb layers in Zr/Nb NMMs: 9.8 to 10.2 GPa. These results indicate that bubbles in Nb are more pressurized than in Zr, with pressure magnitudes closely matching those observed in other NMM systems, such as Cu/V [63]. Since helium molecules fill the bubbles, one may attempt to estimate the He concentration from internal bubble pressure. However, given the high pressures observed, the ideal gas law $c = \frac{p}{kT}$ (where c is the molecule concentration, $p$ is the pressure, k is the Boltzmann constant, and T is the temperature). It is not valid. Instead, we rely on experimental data [105, 106] (obtained at room temperature), which correlate He concentration $C_{He}$ with gas pressure following equation (6).

$$C_{bubble} = 10^{23} A \cdot P^\alpha \tag{6}$$

$C_{bubble}$: Concentration of He in each bubble in He/cm³.

$P$: Pressure in each bubble in GPa.

$A, \alpha$: Parameter to be determined by fit. $A = 0.6227$, $\alpha = 0.4604$

Then, the average He density is given by

$\bar{c} = \frac{\sum_i^N c_i}{N}$ and is plotted in Fig. 8 for all samples. The He concentrations in different regions are as follows:

  i) Pure Zr: 110 to 130 He/nm³

  ii) Pure Nb: 160 to 175 He/nm³

  iii) Zr layers in Zr/Nb NMMs: 80 to 110 He/nm³

  iv) Nb layers in Zr/Nb NMMs: 160 to 190 He/nm³

The He concentration maps in each bubble as a function of depth for pure Zr, pure Nb, and Zr/Nb NMMs are shown in Fig. 8(a-c), respectively. Bubbles are denser in Nb than in Zr, with the He number closely matching those observed in other NMM systems, such as Cu/V [63].

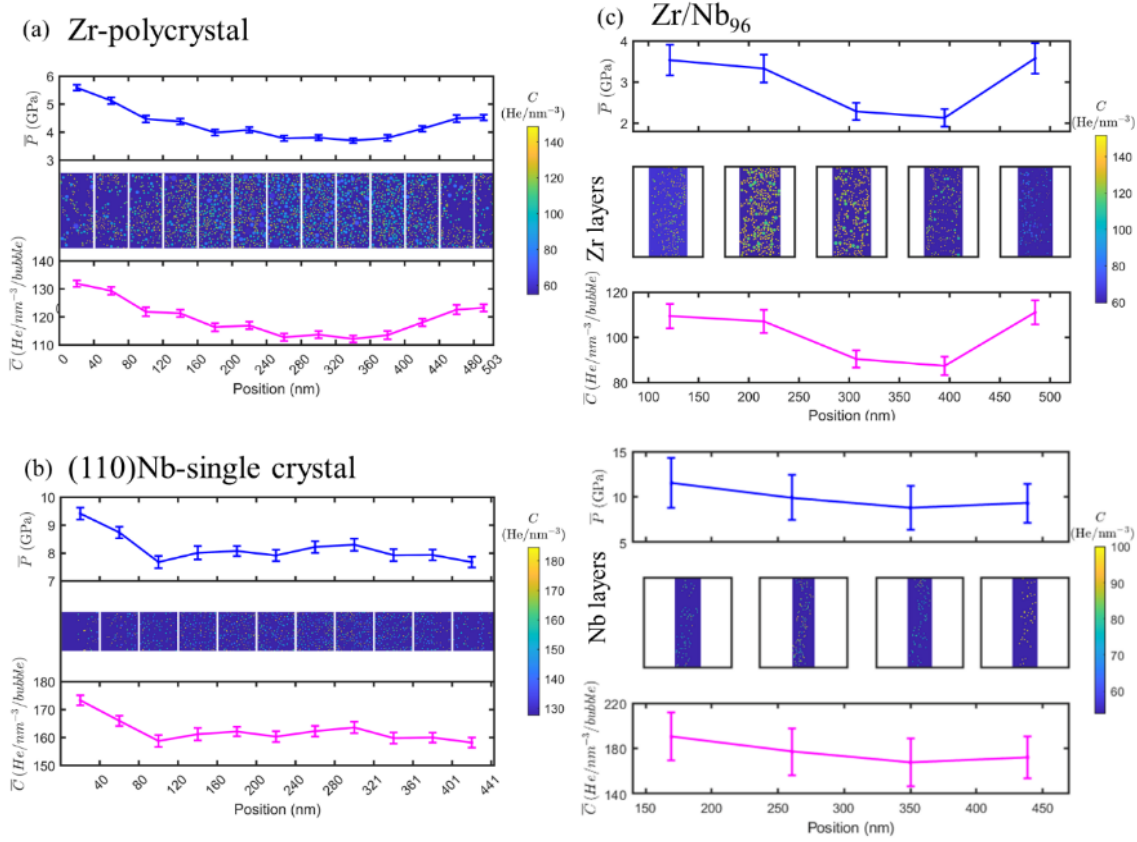

Fig.7. Depth-dependent average pressure and He concentration within bubbles, along with spatially resolved He concentration maps inside each bubble distributed along the ion implantation profile for (a) Zr polycrystal, (b) (110) Nb single crystal, and (c) Zr/Nb$_{96}$ nanoscale metallic multilayers (NMMs) after He irradiation with HF.

The He concentration within individual bubbles, as visualized in the elemental maps of Figure 7, was quantified using the formula: $c_t = \frac{\sum_i^N c_i}{S}$

$c_t$ represents the total He concentration contributing to bubble pressurization, $c_i$ is the He concentration in the $i$-th bubble, $N$ is the total number of bubbles analyzed, and $S$ is the analyzed area. This approach accounts for the cumulative He content distributed across all bubbles within the irradiated region. Figure 8(a-c) illustrates the depth-dependent distribution of He concentrations within bubbles for pure Zr, pure Nb, and Zr/Nb NMMs, respectively. For comparison, the initial implanted He fluence profile, measured via SIMS, is overlaid on each plot. The close alignment between the SIMS profile and the maximum/tail regions of the bubble-derived He concentration confirms the statistical robustness of the experimental measurements. The total He concentration responsible for bubble pressurization (calculated by integrating the blue-shaded areas in Fig. 8) is summarized in the figure captions. Key findings include:

i) Pure Zr: Only 5.2% of the implanted He contributed to bubble formation.

ii) Pure Nb: A significantly higher fraction (17.5%) of the implanted He was retained in bubbles.

iii) Zr/Nb NMMs: 11% of the implanted He participated in bubble formation, with the majority (7%) localized in bubbles within the Zr layers.

These results highlight the critical role of material interfaces in modulating He behavior. In pure Nb (without interfaces), He predominantly resides in bubbles due to favorable vacancy-He interactions. In contrast, the Zr/Nb interfaces in NMMs alter defect dynamics: (i) Vacancies migrate toward interfaces, where they recombine with interstitial atoms, reducing available vacancy sites for He bubble nucleation, and (ii) He is either trapped in small clusters or remains in interstitial positions. This interfacial effect explains the intermediate bubble retention (11%) in NMMs compared to pure Zr (5.2%) and pure Nb (17.5%). The remaining He not contributing to bubbles is attributed to two factors: He trapped in clusters smaller than the TEM detection limit (<0.8 nm in diameter), and or He atoms occupying lattice interstitial sites, which are not resolvable by conventional TEM. The helium-to-vacancy (He/V) ratio within individual bubbles in both Zr and Nb exceeds unity, indicating that the bubbles are overpressurized. This suggests that more helium atoms are present than the number of vacancies available to accommodate them, resulting in significant internal bubble pressure. Such overpressurized bubbles can lead to increased lattice distortion, local swelling, and a higher driving force for dislocation emission. Additional quantitative details and spatial distributions of He/V ratios are presented in Fig. S10.

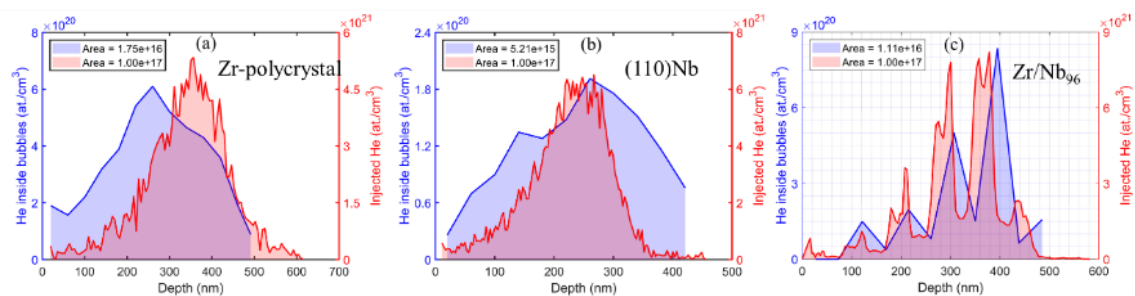

Fig.8. The density inside the bubbles with respect to the irradiated He measured from SIMS of (a) Zr polycrystal, (b) (110)Nb, and (c) Zr/Nb$_{96}$ NMMs.

To verify the presence of He atoms within the bubbles and to experimentally determine the He density inside them, we employed EELS. To the best of our knowledge, such an approach has never been applied to a multilayer system. In metals and alloys, spatially resolved EELS is a powerful technique for determining the He density in individual nanobubbles [107-109]. However, no prior studies have focused on detecting or quantifying He concentration inside bubbles within a multilayered system. For this study, phase mapping was conducted on Zr/Nb$_{96}$ NMMs after high-fluence He irradiation. This analysis confirmed the presence of distinct Zr and Nb layers, as illustrated in Fig. 9a. The HAADF-STEM image reveals black bubble-like structures. Two bubbles were selected for analysis: Bubble 1, with a diameter of 1.54 nm, is situated deep within the Zr layer, far from the interface. Bubble 2, measuring 2.85 nm in diameter, is positioned near the interface (Fig. 9b). EELS spectra characteristic of both the matrix and the bubbles are presented in Fig. 9. The two most prominent signals in the spectra are: The bulk plasmon oscillation of zirconium, which peaks at 20 eV. The zirconium N$_{2,3}$ shell ionization edge, with a maximum at 45 eV (Fig. 9c). Besides Zr signals, extracted EELS spectra reveal the presence of ³He, identified by a sharp peak at ~23 eV, attributed to the ³He K-edge (Fig. 9d, e). The distinct spectral features of ³He and Zr enable a clear distinction between helium nanobubbles and the Zr matrix (Fig. 9). <span style="color:red">After processing, the EELS spectra showing only the He peak are presented in Fig. S11.</span> The He concentration within the bubbles was determined by analyzing the blue shift of the He K-edge relative to its position in a free He atom (21.218 eV) [110, 11]. This shift, ΔE, is related to the helium density ($n_{He}$) through the equation (7).

$$\Delta E = C \; x \; n_{He} \qquad (7)$$

where $n_{He}$ is the He density, and $C$ is a proportionality constant attributed to short-range Pauli repulsion between neighboring electrons. Theoretical and experimental studies on He bubbles in metals have established that C typically ranges between 0.025 and 0.030 eV·nm³ ([104–106]). By applying this method, the measured He 1s-2p energy positions from EELS spectra were: 24.102 eV for Bubble 1 (1.54 nm in diameter), 23.102 eV for Bubble 2 (2.85 nm in diameter). Spatially resolved EELS experiments were performed on bubbles with diameters ranging from 1.54 to 2.85 nm. A representative low-loss spectrum recorded from the center of a bubble (marked by an arrow in Fig. 9b) shows the He K-edge resolved at ~23 eV. From these measurements, the mean helium density was estimated to be 96 He atoms/nm³ for the 1.54 nm bubble and 63 He atoms/nm³ for the 2.85 nm bubble. These values are close to the theoretically calculated densities, which predict 107 He/nm³ and 88 He/nm³, respectively (Fig. 7).

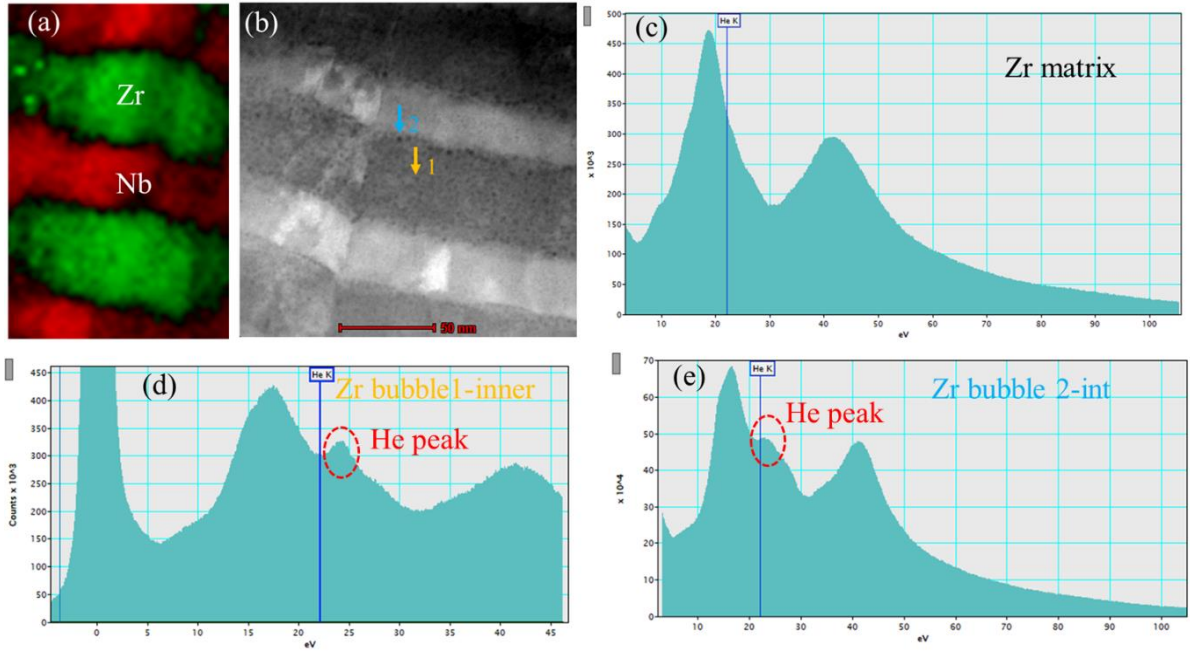

Fig.9. (a) ASTAR phase mapping, (b) HAADF STEM of He irradiated Zr/Nb$_{96}$ with HF, EEL spectrum recorded from (c) the Zr matrix, (d) from the center of a He bubble in the inner region of Zr layer, (e) from the center of a He bubble in the interface Zr side. Dashed red circles in d, e, highlight He peaks from bubble 1 (inner Zr layer), and from bubble 2 (at the interface), respectively.

4.2 Analysis of irradiation Hardening

For weak obstacles, such as He bubbles [63], a hardening relationship developed by Friedel-Kroupa-Hirsch FKH is applied to describe the dependence of radiation hardening on He bubbles:

$$\Delta \sigma = \frac{1}{8} M \mu b d N_{He}^{2/3} \qquad (8)$$

where M is Taylor factor, 3.06, $\mu$ is the shear modulus, b is the Burgers vector, and d is the diameter of bubbles. By using b=0.286 nm, 0.32 nm for Nb and Zr, respectively, $\mu$ an average shear modulus of Nb of 37.5 GPa, and for Zr of 33 GPa, $d$ from Figs 2, 3, and 4.

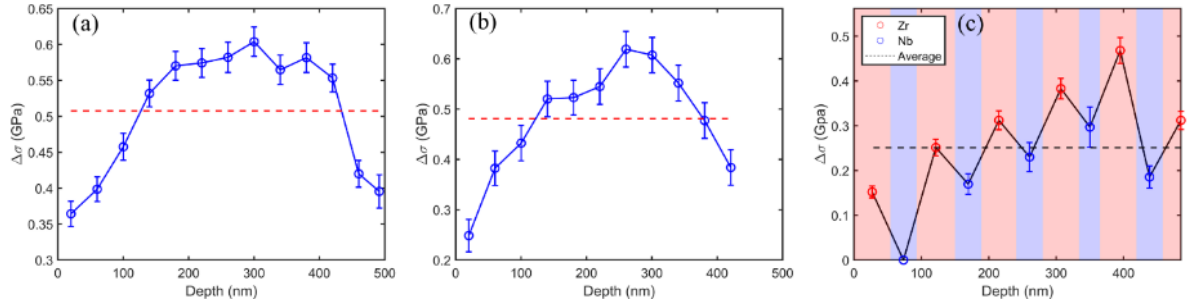

Fig.10. dependence of radiation hardening on He bubbles from FKH model vs penetration depth: (a) Zr polycrystal, (b) (110)Nb, and (c) Zr/Nb$_{96}$ NMMs after He ion irradiation with HF.

The average increase in yield strength from the FKH model is calculated to be 0.51, 0.49, and 0.27 GPa for Zr, Nb, and Zr/Nb$_{96}$, respectively, as shown in Fig.10. Correspondingly, hardness enhancements of 1.9 GPa in Zr, Nb, and 1.7 GPa in Zr/Nb$_{96}$ are obtained, assuming that hardening from the bubbles is one-third of the measured hardness. Thus, radiation hardening due to He bubbles is very small compared to experimental values. Other factors, such as He-vacancy clusters (less than 0.5 nm) and interstitial loops, may also contribute to hardening. Since such He clusters are undetectable by TEM, DBS-PALS was employed to confirm their presence.

The detection of defects using positrons is based on the principle that positrons can be trapped by lattice defects [87, 88]. This occurs because positrons are strongly repelled by the positively charged ion cores in the lattice; therefore, the absence of a positive charge at a vacancy creates an attractive potential that can trap positrons. A typical trapping site binds the positron with an energy on the order of 1 eV, sufficient to keep the thermalized positron trapped until annihilation occurs. From Fig.11 (a-b): Zr-Nb$_{96}$ vs. He-irradiated Zr-Nb$_{96}$, the S parameter (low-momentum electron fraction), S increases significantly near the surface (0–20 nm) and remains constant. The increase in S in irradiated samples indicates the formation of open volume defects (vacancies, small voids) near the surface due to He implantation. The increase is rather small, likely due to positron saturation trapping effects (high number of pristine defects) and/or a large resilience of the material to the ion damage. The saturation suggests that defect distribution stabilizes with depth. Irradiated samples (red circles) show slightly higher S than virgin ones (black squares), confirming damage due to He ions. The W parameter (high-momentum electron fraction): W decreases after irradiation. A decrease in W supports the increase in low-momentum electron environments, confirming the formation of open-volume defects rather than heavier impurity atoms.

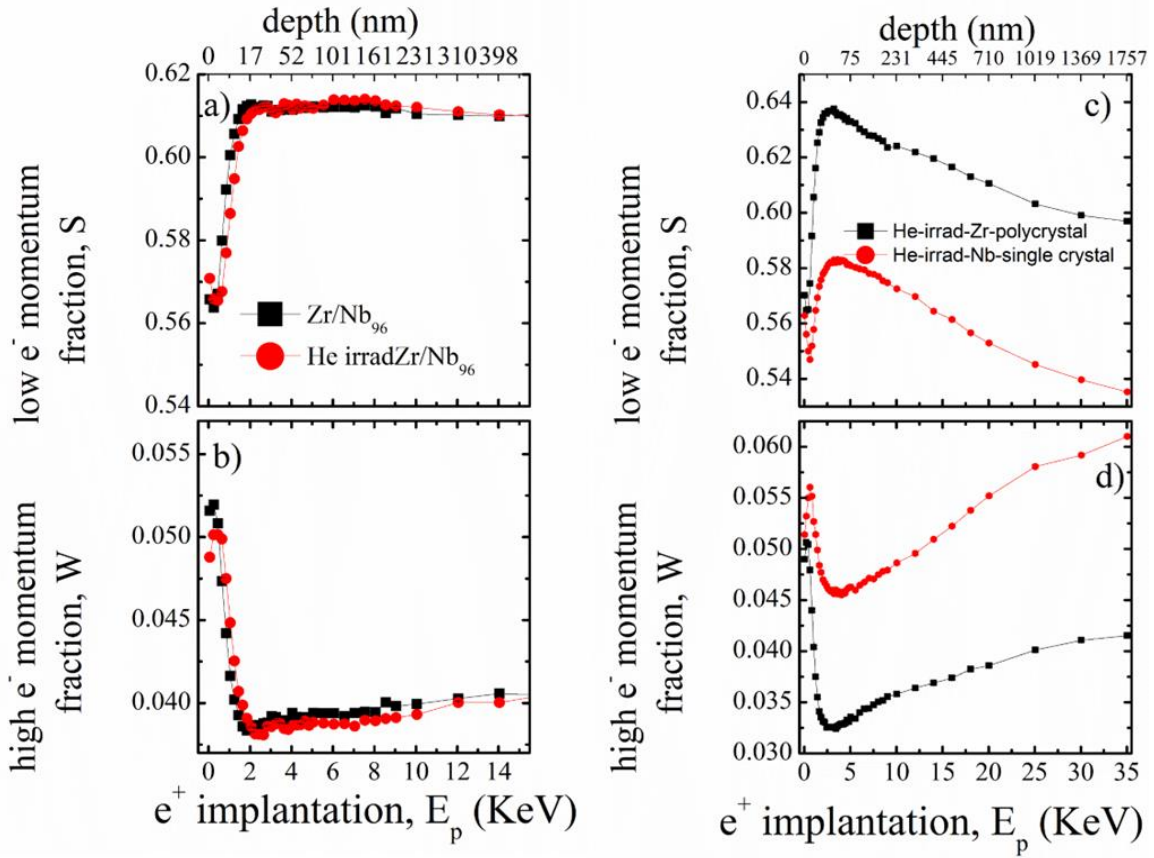

Fig.11. Doppler Broadening Spectroscopy (DBS) analysis of He-irradiated and unirradiated Zr–Nb$_{96}$ (left panel) and comparison between He-irradiated Nb single crystal and Zr polycrystal (right panel): (a, d) Low momentum S-parameter as a function of positron implantation energy and depth, indicating the concentration of open volume defects; (b, e) High momentum W-parameter representing annihilation with core electrons, inversely correlated with vacancy-type defects.

The comparison between Nb single crystals and Zr polycrystals (Fig. 11c, d) shows lower defect concentration in Nb, which may stem from lower helium mobility and reduced recombination efficiency in the single-crystalline lattice, hence increased He retention in defects and their agglomerations. In contrast, polycrystalline Zr shows larger defect concentration and saturation, likely near grain boundaries, which act as effective sinks.

The bulk Nb-single crystal and Zr-polycrystalline sample irradiated with He were measured with PALS, and the results are presented in Fig. 12.

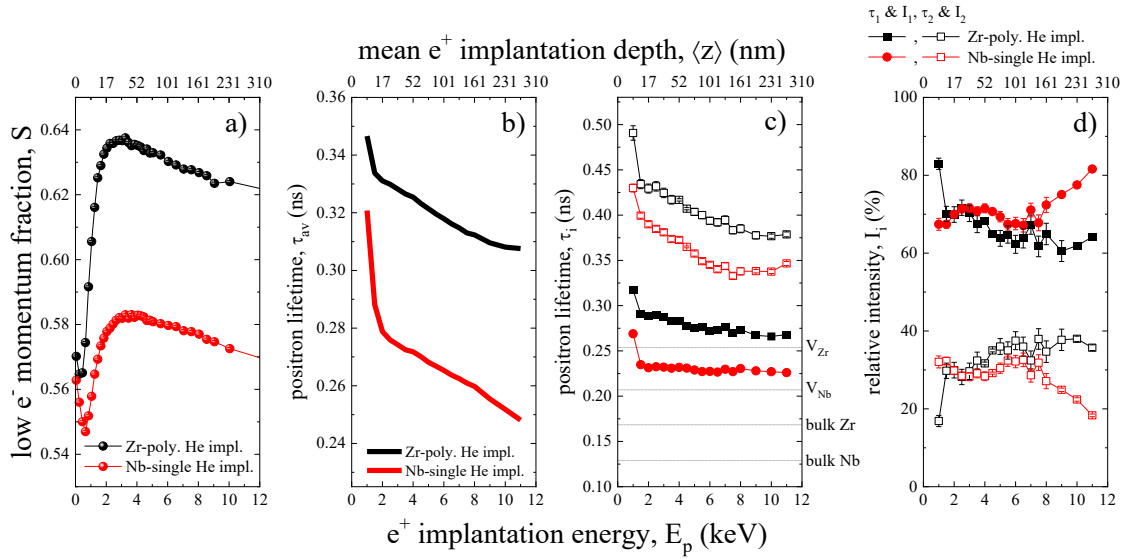

Fig. 12. DBS of He-irradiated bulk Nb and Zr samples (a), average positron lifetime (b), positron lifetime components $\tau_1$ and $\tau_2$ (c), and their relative intensities $I_1$ and $I_2$ (d). The dotted lines in (c) represent calculated positron trapping states from literature [112].

DBS (Fig. 12a) and average positron lifetime (Fig. 12b) $\tau_{av}$ ($\tau_{av} = \Sigma\, \tau_i \cdot I_i$) are in good agreement, as a larger overall defect size is expected in Zr. The majority defect ($I_1 \sim 70\%$) consists of small vacancy agglomerations, possibly decorated with He, since the measured lifetime $\tau_1$ is slightly larger than a respective Zr or Nb monovacancy. The other defects $\tau_2$, contributing with intensity of $I_1 \sim 30\%$ represent larger vacancy clusters. Both $\tau_1$ and $\tau_2$ are larger for Zr, which is a consequence of the crystal structure, but could be related to higher He retention in Nb. No void-related components were found, which suggests overall large He content in the bulk samples or no bubble formation at all. The combined PAS results offer a powerful insight into the defect evolution in Zr–Nb$_{96}$ under helium irradiation, particularly for detecting nanometer- and sub-nanometer-scale open-volume defects, which are beyond the detection limit of traditional imaging techniques such as TEM.

Positron lifetime measurements (Fig. 13) confirm the presence of multiple defect types. The $\tau_1$ and $\tau_2$ components correspond to bulk annihilation, mono-vacancies and small vacancy clusters [112, 113], respectively, while $\tau_3$ and $\tau_4$, present mostly in irradiated samples, indicate the formation of large-scale voids. Notably, the defect size ($d_3$ and $d_4$) and relative intensity ($I_3$, $I_4$) increase significantly with positron implantation energy and irradiation fluence. The most severe damage is observed in Zr–Nb$_{96}$ HF, where τ$_4$ and I$_4$ rise steeply beyond 6 keV, signifying the growth of irradiation-induced voids at deeper regions.

Together, the DBS and PALS analyses confirm that He irradiation leads to a depth-dependent accumulation of open-volume defects in Zr-Nb$_{96}$, with severity increasing at higher fluences. These results have critical implications for understanding irradiation swelling, mechanical hardening, and microstructural evolution in advanced nuclear cladding materials.

Depending on imaging contrast and sample preparation, TEM is generally limited to detecting pores or bubbles larger than ~1-2 nm. However, the PALS technique used here provides sensitivity to the sub-nanometer range (~0.2-1 nm), making it especially suitable for detecting early-stage void nucleation and vacancy clustering [114]. The lifetime components τ$_3$ and τ$_4$, and their associated pore sizes d$_3$ and d$_4$, clearly show the formation of voids with diameters ranging from ~0.5 to >2 nm in irradiated samples. These voids are too small and sparse to be resolved via TEM but have a measurable influence on mechanical properties.

In the He-irradiated Zr–Nb$_{96}$ samples, increasing positron implantation energy reveals a growth in τ$_4$ and d$_4$, particularly in the HF condition. Pore size increases progressively with depth, confirming depth-dependent void swelling likely due to defect accumulation and bubble coalescence. The DBS results for He-irradiated pure Nb (single crystal) and pure Zr (polycrystal) demonstrate differing behaviors in defect accumulation:

- He-irradiated Nb single crystal shows a lower S-parameter and higher W-parameter near the surface than Zr, indicating a lower concentration of vacancy-type defects and voids as they could be filled with He. The low mobility of defects and He in Nb, due to its single-crystal nature and lack of grain boundaries, may contribute to the retention of helium-vacancy complexes and at larger defect clusters.

- In contrast, He-irradiated Zr polycrystals display a larger S-parameter across the overall depth, suggesting a large number of defects but an efficient He out transport due to the presence of grain boundaries, which serve as efficient sinks for radiation-induced point defects as well.

- The Zr/Nb$_{96}$ multilayer structure (alternating layers of Zr and Nb, 96 nm period) shows an intermediate behavior. Compared to its monolithic counterparts, the multilayer benefits from interface-assisted defect recombination, resulting in overall lower contribution (I$_2$~20%) of lifetimes related to vacancy agglomerations (τ$_2$). In addition, $\tau_2$ reduces with He implantation, which could be a fingerprint of its retention or indeed interface-assisted defect recombination.

These results underline the importance of multilayer architectures in managing irradiation-induced defects. The Zr/Nb$_{96}$ structure not only delays void swelling compared to pure Nb or Zr but also suppresses large pore formation, thanks to the high density of interfaces that facilitate point defect recombination. The absence of visible bubbles in TEM, despite clear swelling (as indicated by PALS), further emphasizes the critical role of advanced positron-based spectroscopies in detecting early-stage damage in radiation environments.

The synergy between helium irradiation and microstructural design strongly influences pore formation and radiation damage evolution. PALS and DBS provide essential insight into nano-scale defects, invisible to TEM, offering a more complete understanding of material performance in extreme nuclear environments.

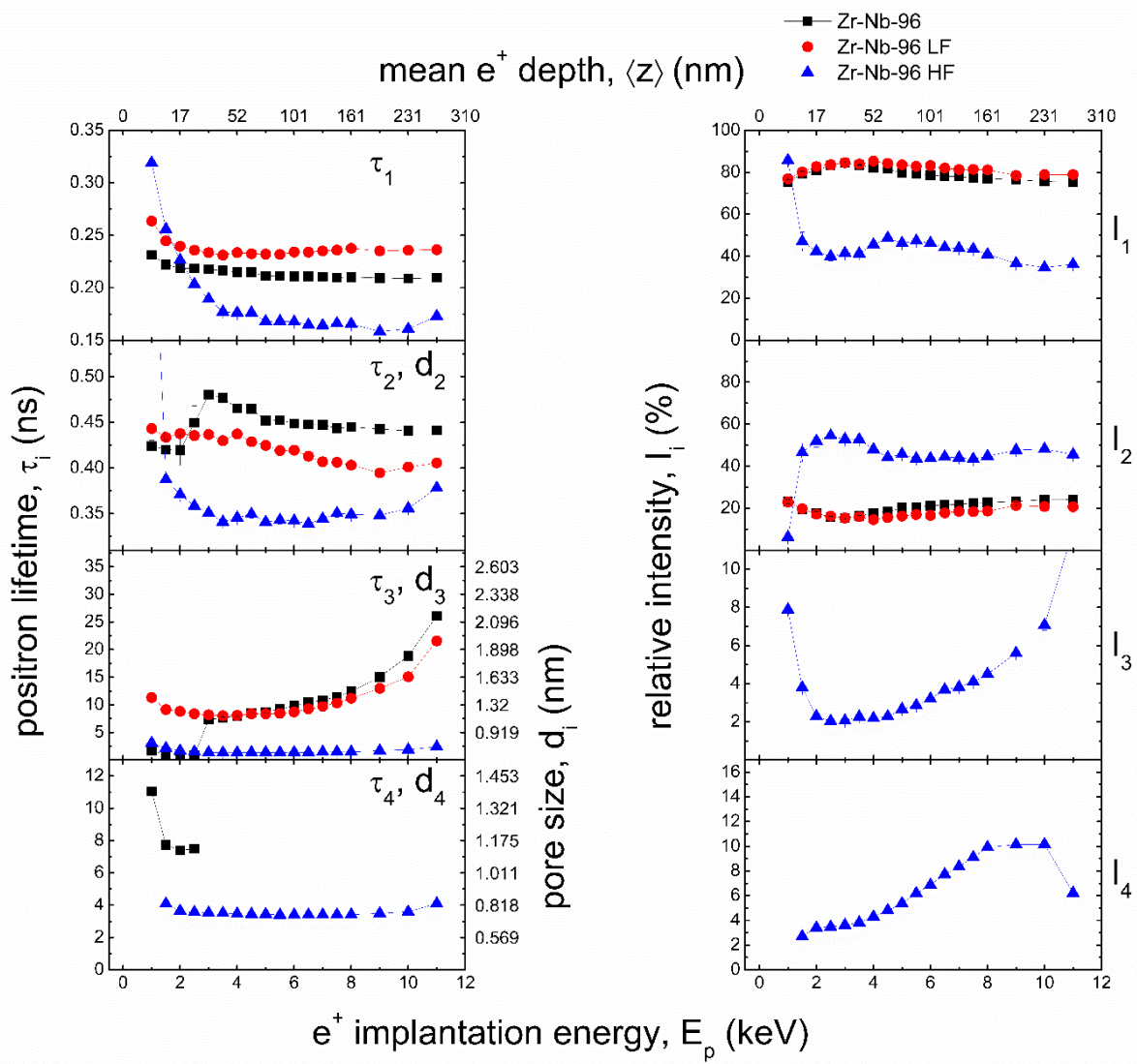

Fig.13. Positron Annihilation Lifetime Spectroscopy (PALS) results for virgin and He-irradiated Zr/Nb$_{96}$ alloys at different fluences: (left panel) Lifetime components $\tau_1$–$\tau_4$ and corresponding pore sizes $d_1$–$d_4$ reflect different vacancy clusters and void sizes; (right panel) Relative intensities $I_1$–$I_4$ indicate the fractional contribution of each defect type. Higher $\tau_3$, $\tau_4$, and increasing intensities $I_3$ and $I_4$ with positron implantation energy suggest the evolution of larger voids at greater depths in high-fluence irradiated samples (Zr-Nb$_{96}$ HF). This confirms depth-dependent vacancy clustering and void swelling due to He irradiation.

4.3 Interface-affected zones in nanolaminates through Cascade/DFT simulations

A detailed TEM analysis of the Zr/Nb nanolayered material highlights the critical role of interfaces in governing defect dynamics with three key insights. First, polycrystalline Zr exhibits larger He bubbles than (110)-oriented Nb. Second, within Zr/Nb NMMs, the density of He bubbles is higher in the Zr layers than in the Nb layers. Third, BDZs consistently form on the Nb side of the Zr/Nb interface. The mechanisms driving bubble alignment in Zr and bubble suppression in Nb remain a subject of investigation. Given the modest lattice mismatch (~2.17%) between Zr and Nb, epitaxial growth at the interface is feasible, likely minimizing strain and misfit dislocations. To further probe the behavior of point defects and helium near and away from the interface, we performed several simulations at the nanoscale.

A plausible mechanism for these observations involves differential damage recovery behaviors in the constituent layers: Nb may recover via efficient defect recombination, whereas Zr remains heavily damaged, retaining large vacancy clusters. To investigate this hypothesis and elucidate damage evolution in both bulk and interface systems, classical molecular dynamics (MD) simulations were conducted using the primary knock-on atom (PKA) methodology. In this approach, a designated atom is imparted with an initial kinetic energy (3 MeV in this study), initiating a collision cascade and associated defect generation.

Figures 14a and 14b schematically depict this process within an interfacial system. The PKA, indicated in Figure 14a, initiates damage that propagates and is visualized in white in Figure 14b. Over time, the damaged region diminishes due to recombination processes. After approximately 60 ps, the system reaches a dynamic equilibrium, as evidenced by the stabilization of the total defect count over the final 30 ps. At this point, either complete healing occurs, or a small number of residual defects persist.

Figure 14c presents the temporal evolution of dislocations in bulk Nb, bulk Zr, and the interface system. For interface cases, the PKA was placed ~10 Å from the interface, and simulations were conducted with knock directions both toward and away from the interface. The results were then averaged. These data indicate that Zr is intrinsically more susceptible to irradiation-induced defect formation than Nb, consistent with the higher defect formation energies in Nb [1115], which render atomic displacements energetically more costly.

Interestingly, the Zr layer within the interfacial system sustains less initial damage (see "max" in Figure 14c) compared to bulk Zr. This behavior is attributed to the directional nature of the cascade: PKAs knocked toward the interface in the Zr layer impart some of their damage to the adjacent Nb, thereby reducing net defect formation in Zr. Conversely, the Nb layer in the interface system reaches a higher maximum damage than its bulk counterpart, as cascades directed toward the interface induce additional damage in the neighboring Zr layer which is easier to displace. Following the cascade, recombination processes initiate rapidly, reducing defect concentrations to ~10% of their peak values within a few picoseconds. Although the recombination rate is comparable in both bulk and interface systems, Nb exhibits more rapid recombination kinetics than Zr. Notably, the presence of the interface does not significantly influence the recombination rate, but it does affect the extent of healing (Figure 14d). Figure 14d also illustrates the final number of residual defects after 60 ps as a function of the PKA's initial location. Data are averaged over knock directions toward and away from the interface. The results indicate that both vacancy and interstitial defect populations are higher in bulk systems (edges of the plot) compared to the interface region, suggesting that interfaces enhance defect recombination and facilitate healing [63]. Moreover, the proximity of the PKA to the interface correlates inversely with the residual defect count. Interestingly, Zr heals more effectively in the presence of an interface compared to Nb, as the Nb side retains more residual defects at steady state.

From the PKA simulations, four primary conclusions can be drawn, although these alone do not fully account for the experimental observations: i) Zr is more susceptible to defect generation under irradiation; ii) Nb exhibits faster recombination kinetics; iii) Zr shows greater long-term healing efficiency; iv) while interfaces do not accelerate recombination rates, they do enhance the net healing effect.

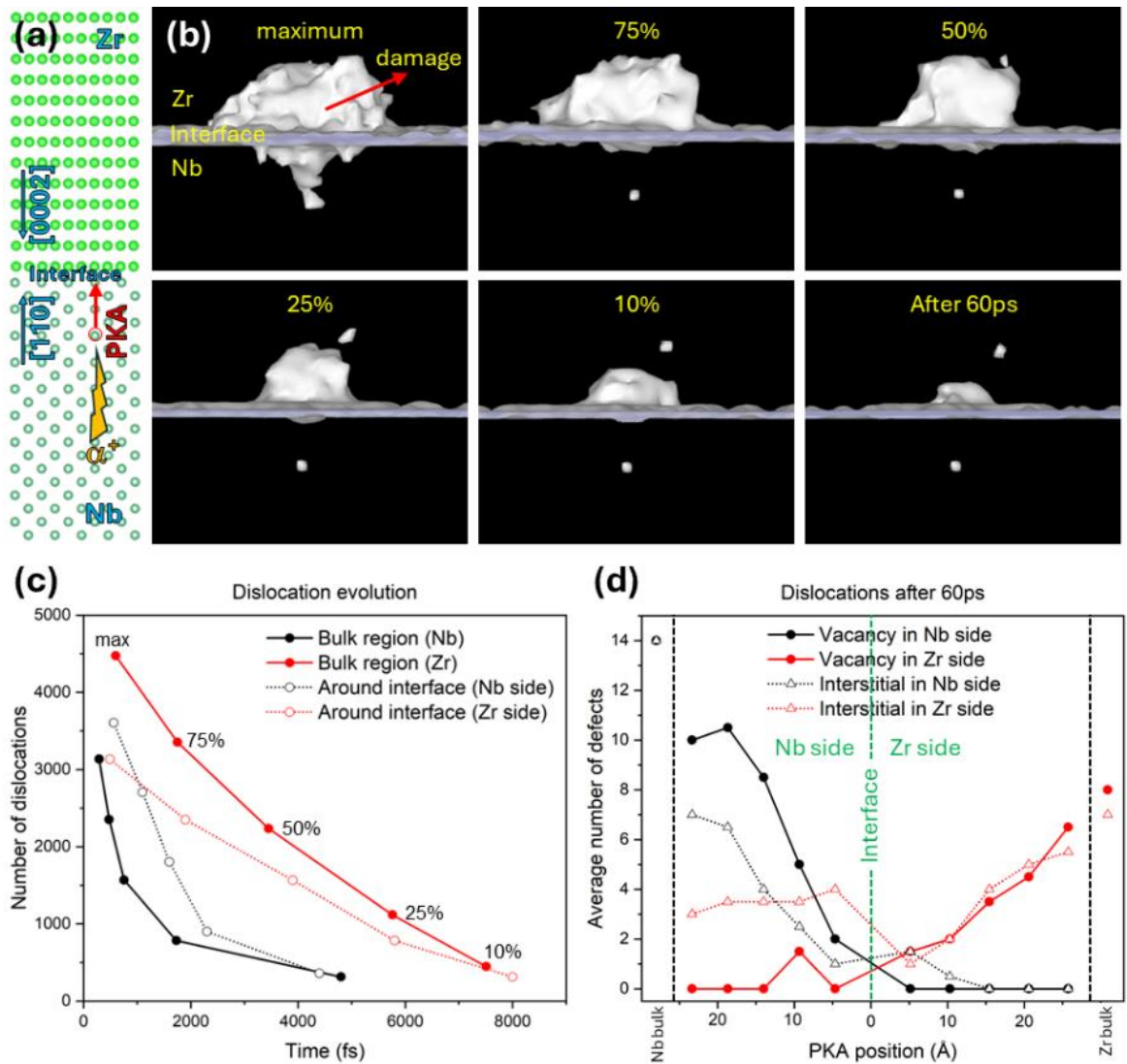

Fig. 14. a) Schematic of the (110)Nb/(0002)Zr interface system with a PKA directed toward the interface. b) Damage evolution (white regions) during cascade events, showing the initial damage peak and subsequent reduction over time. c) Temporal evolution of dislocations in bulk and interface systems; for interface cases, data are averaged for PKAs directed both toward and away from the interface. d) Final defect count after 60 ps as a function of initial PKA position.

To further assess steady-state defect distributions, we implemented a one-dimensional reaction-diffusion model following the formalism of Demkowicz et al. [75]. The model parameters are listed in Table 1, and the results are shown in Fig. 15. This approach couples point-defect production, recombination, and diffusion within the Zr and Nb layers under continuous irradiation. The calculated steady-state profiles exhibit a pronounced asymmetry:

defect concentrations are significantly higher on the Nb side than on the Zr side, while defect concentrations decrease progressively toward the interface, reflecting its strong sink effect. The higher steady-state concentration in Nb arises from its lower effective defect diffusivity, which limits migration to the interface for recombination. These predictions are consistent with MD simulations showing fewer defects adjacent to the interface and relatively less damage in Zr layers. However, both the model and the MD simulations, taken alone, predict a greater tendency for bubble formation on the Nb side due to the higher defect concentration, opposite to what is observed experimentally. This reinforces that He transport and clustering dynamics must be explicitly included to fully explain the observed BDZs.

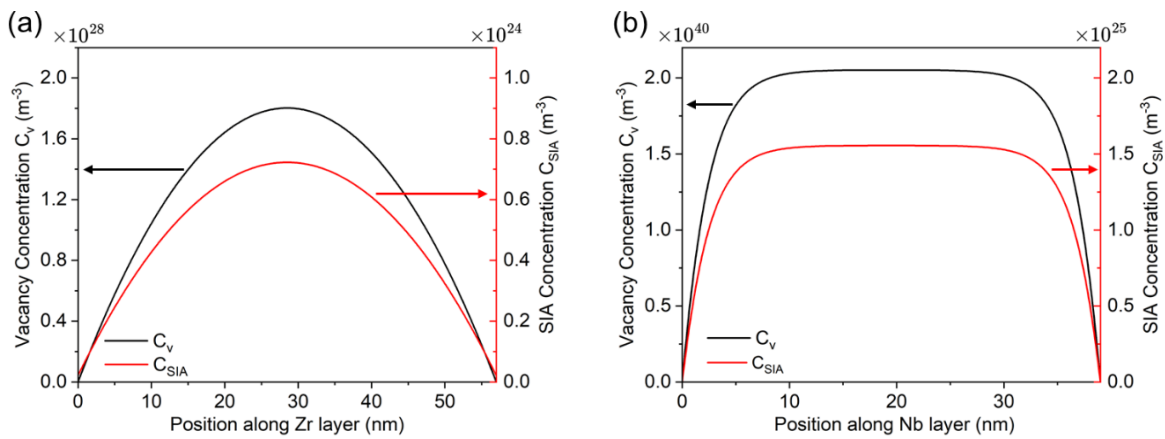

Fig. 15. Vacancy (black line) and self-interstitial (red line) atom concentrations within the Zr (a) and Nb (b) layers, calculated using a one-dimensional reaction–diffusion model at steady state and room temperature. The interface positions are represented by the plot edges. The profiles reveal higher defect concentrations on the Nb side compared to the Zr side, with concentrations decreasing toward the interface, consistent with its strong sink effect.

The experimental observation of larger and more numerous He bubbles in Zr layers suggests post-cascade growth through vacancy and He atom aggregation. While MD simulations indicate that the irradiated lattice heals significantly through recombination, some He atoms likely become trapped before complete defect recovery. These trapped He atoms may subsequently nucleate and grow into bubbles by capturing additional He atoms or vacancies. To evaluate this hypothesis, we estimated the room-temperature diffusivities of vacancies (V), self-interstitial atoms (SIAs), and He atoms in Zr and Nb. Ab initio molecular dynamics (AIMD) simulations were employed to assess He mobility, whereas classical MD simulations were used for V and SIA diffusivities due to their inherently lower mobility and the impracticality of converging such slow processes via AIMD.

All simulations were performed at elevated temperatures (500–1700 K) to accelerate atomic mobility. Mean squared displacements (MSDs) were computed from the atomic trajectories and diffusion coefficients (D) extracted using the Einstein relation:

$$D = \lim_{t \to \infty} \frac{\langle \Delta r^2(t) \rangle}{6t} \tag{9}$$

where $\langle \Delta r^2(t) \rangle$ is the time-averaged mean squared displacement. [116] The temperature-dependent diffusivities were then fitted to an Arrhenius-type[116] expression:

$$D(T) = D_0 e^{-\frac{E_a}{k_B T}} \tag{10}$$

where $D_0$ is the pre-exponential factor, $E_a$ is the activation energy for diffusion, $k_B$ is Boltzmann's constant, and $T$ is the temperature.

Extrapolated room temperature diffusivities (Fig. 16) indicate that He atoms are significantly more mobile than both vacancies and SIAs in both materials. The estimated room temperature diffusivities are summarized in Table 1. These results confirm that He atoms can readily diffuse and become trapped by vacancies before recombination, supporting the formation and growth of He bubbles.

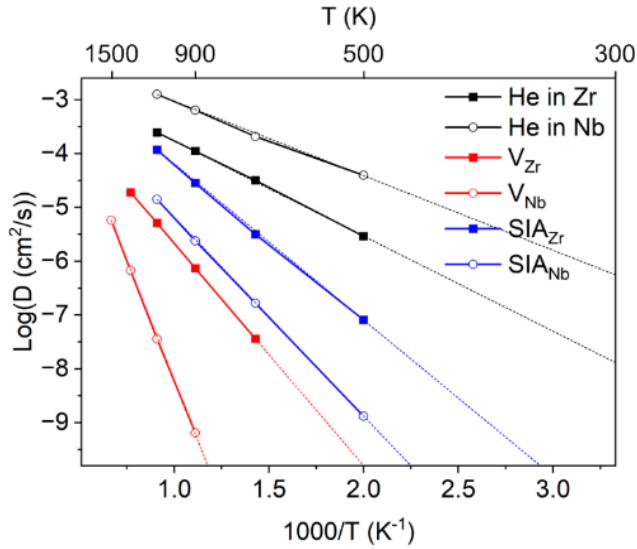

Fig. 16. Diffusivity of He atoms and vacancies in bulk Zr and Nb as predicted by molecular dynamics simulations.

Table 1: Parameters for Zr and Nb used in the one-dimensional reaction–diffusion model, along with helium diffusivities estimated by ab initio molecular dynamics simulations at room temperature.

| Quantity | Zr | Nb |
| --- | --- | --- |
| $E_f(V_1)$ (eV) | 2.03 | 2.62 |
| $E_f(SIA)$ (eV) | 3.1 | 4.0 |
| $E_m(V_1)$ (eV) | 0.82 | 1.76 |
| $E_m(SIA)$ (eV) | 0.57 | 0.73 |
| Number of sites surrounding a vacancy | 12 | 8 |
| $K_0$ (cm$^{-3}$s$^{-1}$) | \multicolumn{2}{c|}{$1.0 \times 10^{19}$} |
| $D_V$ (cm$^2$/s) | $5.0 \times 10^{-16}$ | $8.3 \times 10^{-30}$ |
| $D_{SIA}$ (cm$^2$/s) | $1.1 \times 10^{-11}$ | $1.6 \times 10^{-14}$ |
| $D_{He}$ (cm$^2$/s) | $1.3 \times 10^{-8}$ | $5.5 \times 10^{-7}$ |

To gain further insight into vacancy clustering, we computed vacancy cluster formation and trapping energies using the expressions:

$$E_f = E(V_n) - (E_{perfect} + n \times \mu) \text{ and } E_T = E(V_n) - E(V_{n-1} + V_1) \qquad (11)$$

where $E(V_n)$ is the total energy of the system with a cluster of $n$ vacancies, $E_{perfect}$ is the total energy of a defect-free system, $\mu$ is the atomic chemical potential (-10.21 eV for Nb and -8.52 eV for Zr) defined as the total energy per atom for bulk structures, and $E(V_{n-1} + V_1)$ is the total energy of the system with a cluster of $n-1$ vacancies and a single vacancy away from the cluster (see Fig. 17a). As shown in Fig. 17b, vacancy cluster formation energy increases almost linearly with cluster size. The formation of clusters requires more energy in Nb compared to Zr, which is consistent with the initial damage observed in PKA simulations. The trapping energy, on the other hand, remains constant at about -0.45 eV and -0.25 eV for Nb and Zr systems, respectively (Fig. 17c). The negative and consistent values with cluster growth indicate that larger clusters are energetically favorable compared to smaller, dispersed ones.

Once a He is trapped inside a vacancy or vacancy cluster, the V-He bubble might grow by capturing other He atoms and/or vacancies in the vicinity to become even larger. To explore He-vacancy cluster growth, we calculated He trapping energy for an already existing single vacancy-He cluster. We define the trapping energy for a He atom as:

$$E_T = E(V_1 He_n) - E(V_1 He_{n-1} + He_I) \qquad (12)$$

where $E(V_1 He_n)$ is the total energy of the system with a single vacancy occupied by $n$ He atoms while $E(V_1 He_{n-1} + He_I)$ is the total energy of the system with a single vacancy occupied by $n-1$ He atoms and a distant single He atom at an interstitial site (see Fig. 17d). As shown in Fig. 16e, Nb and Zr can trap up to 3 and 6 He atoms, respectively, in a single vacancy before trapping becomes energetically unfavorable. Corresponding He-vacancy cluster size of 15 Å$^3$ for Nb and 35 Å$^3$ for Zr (Fig. 17f) are consistent with the lower bulk and Young's moduli of Zr compared to Nb[109] in line with the larger He bubbles observed experimentally in Zr layers.

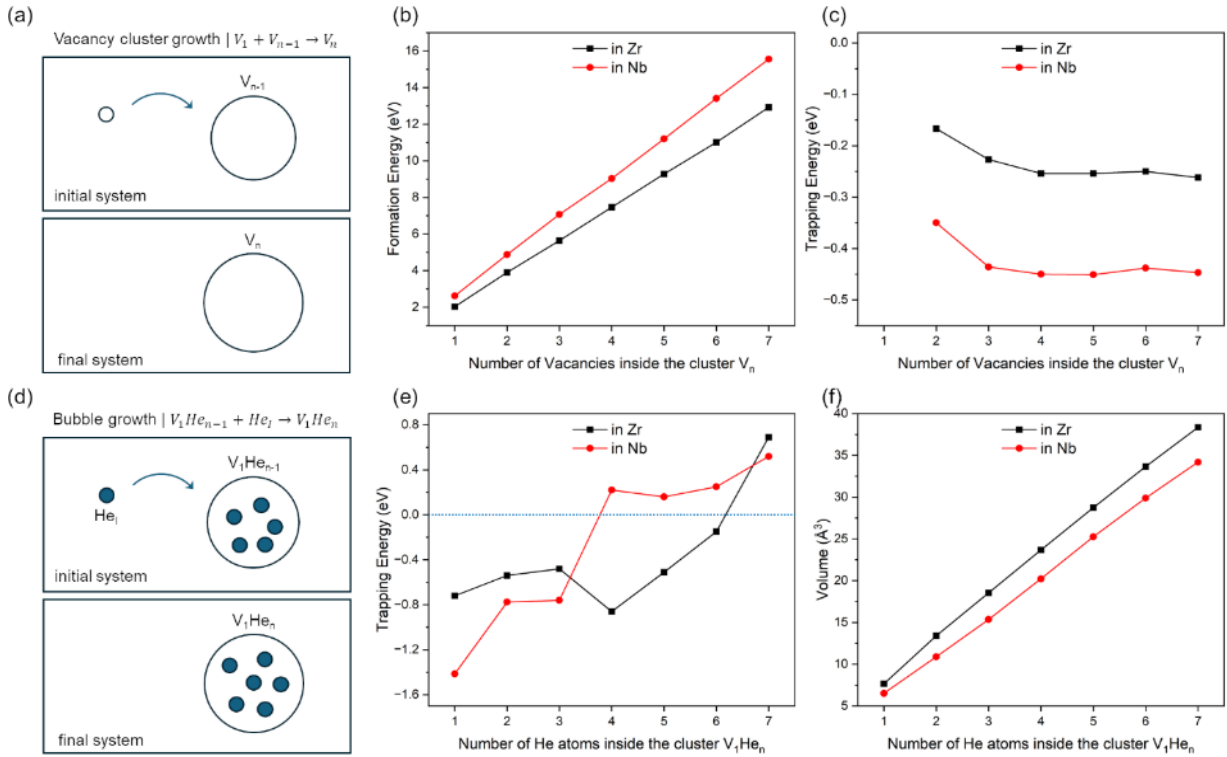

Fig.17. (a) Schematic for the vacancy cluster growth. (b) Vacancy cluster formation energy as a function of size. (c) Trapping energy of a single vacancy by a vacancy cluster. (d) Schematic for the bubble growth. (e) Trapping energy of a He atom by the bubble. (f) Volume change of the bubble as a function of the number of captured He atoms.

The experimental observations consistently reveal the presence of BDZs on the Nb side of the Nb/Zr interface. However, the DFT simulations presented thus far have been limited to bulk structures and do not account for interfacial effects. To explicitly investigate the role of the interface, we performed NEB calculations to evaluate the migration energetics of a single vacancy and a helium atom toward a vacancy cluster (V$_4$) in bulk, and toward the interface in interfacial models, as illustrated in Fig.18a. In both systems, the center of the vacancy cluster or the center of the interface is designated as layer 0, with surrounding atomic layers indexed by increasing integers to indicate the diffusion pathway.

Figure 18b presents the migration energetics of a single vacancy toward the cluster or interface, and the magnitude of the energy barrier is summarized in Table S1. In all cases, the total energy of the system decreases as the vacancy approaches the cluster or interface, indicating that such migration is thermodynamically favorable. Notably, upon reaching the second layer from the interface/cluster center, the vacancy spontaneously migrates to the first layer, suggesting a strong energetic driving force for localization near the interface or within the cluster. The calculated migration energy barriers ($E_m$) for vacancies are consistently higher in Zr than in Nb, in both bulk and interface systems. This disparity implies that vacancies near the interface diffuse more readily from the Nb side, contributing to the accumulation of vacancies at the interface and the formation of BDZs on the Nb side.

The corresponding energetics for He atom migration are shown in Fig. 18c. On Nb side, there are several stable interstitial sites for He atoms to reside, which results in the rapid fluctuations in energy, while in the Zr side, He atom migration requires longer jumps from one interstitial site to another, which is in agreement with our previous work [117]. Similar to vacancies, He atoms also exhibit a preference to migrate toward the vacancy cluster or interface. In the cluster case, He atoms settle at the center of the $V_4$ cluster. At the interface, however, He atoms preferentially localize on the first Zr layer, in agreement with earlier findings [115, 117]. Interestingly, He atoms in Nb can migrate toward the interface or cluster from as far as two layers away without encountering an energy barrier, whereas in Zr, a finite barrier persists even within the first few layers. This suggests a more facile He migration toward the interface from the Nb side. Furthermore, the migration barriers for He atoms toward the interface are significantly lower than those toward the vacancy cluster for both Nb and Zr, indicating the interface acts as a stronger sink for He. Around the interface, the He migration barrier is notably lower on the Nb side compared to the Zr side, further supporting the enhanced He diffusion towards the interface in the Nb side. This asymmetry in transport behavior underscores the role of the interface in facilitating the accumulation of defects from the Nb side, leading to a depletion of both He atoms and vacancies in the adjacent Nb layers-a phenomenon that manifests as the experimentally observed BDZs.

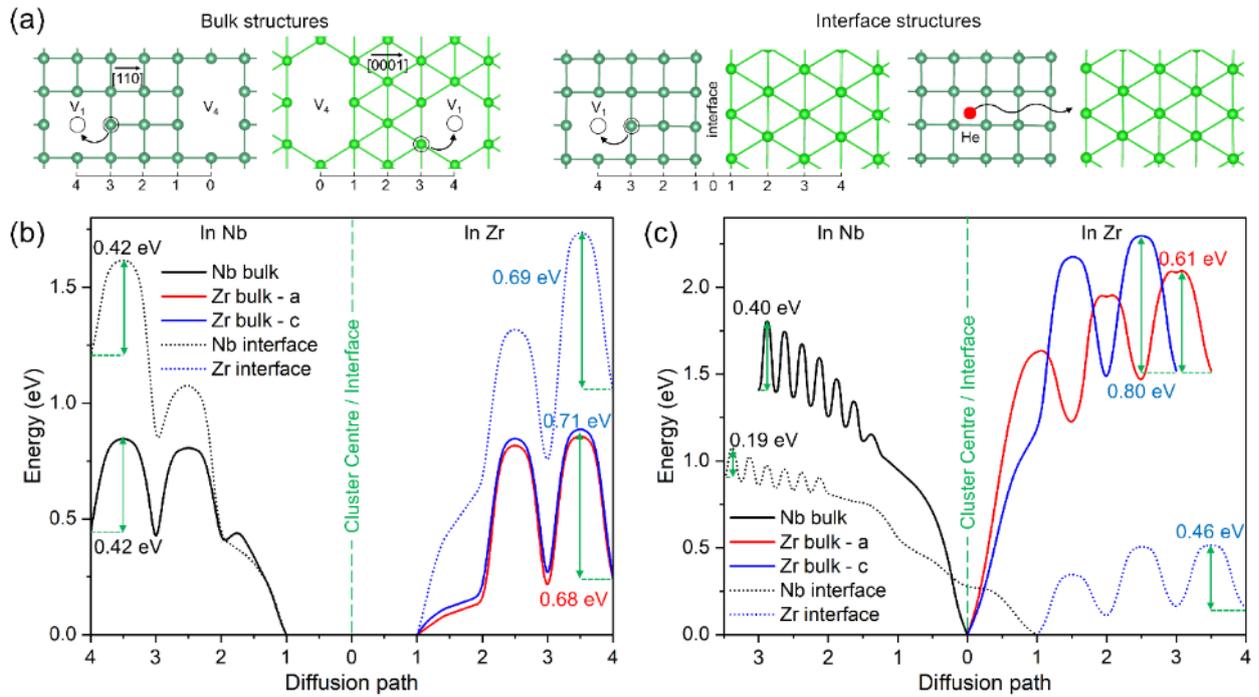

Fig.18. (a) Schematic for the diffusion of a single vacancy or a He atom towards the cluster of $V_4$ and the interface. (b) Energetics for the vacancy migration in Nb ([110] direction), Zr (a/c directions) towards the cluster of $V_4$, and towards the interface. (c) Energetics for the He atom migration in Nb ([110] direction), Zr (a/c directions) towards the cluster of $V_4$, and the interface.

While the NEB calculations demonstrate that asymmetric migration barriers are the primary driving force for the formation of BDZs, the atomic structure of the interface can also influence local helium trapping. Prior work on Cu/Nb interfaces [45] showed that platelets form preferentially on the Cu side due to its lower vacancy formation energy, which, with continued He and vacancy accumulation, evolve into larger bubbles. By analogy, the lower vacancy formation energy of Zr compared to Nb suggests that interfacial He platelets or bubbles are more likely to nucleate on the Zr side of the Zr/Nb interface. Our earlier work [115] further indicates that Nb vacancies at the interface can be healed by capturing a Zr atom from the adjacent layer, effectively shifting vacancy sites into Zr. These mechanisms are consistent with the experimentally observed He bubbles in Zr. However, they do not explain the absence of bubbles in Nb, where bubble nucleation would otherwise be expected. The NEB results resolve this asymmetry by showing that the lower migration barriers for both vacancies and He atoms in Nb drive a flux of defects toward the interface, supplying helium for Zr-side bubbles while depleting the adjacent Nb layers. Thus, the experimentally observed

<span style="color:red">asymmetric He distribution originates from migration-barrier-driven defect flux, with vacancy energetics providing a consistent explanation for Zr-side nucleation.</span>

## 5. Conclusions

This study reveals that Zr/Nb nanolaminates with BCC/HCP interfaces mitigate helium irradiation damage through asymmetric defect dynamics. Under helium fluences ($1\times10^{16}$–$1\times10^{17}$ He/cm$^2$), Zr layers host larger bubbles (~1.5–2.8 nm, 1.2% swelling), while Nb layers form bubble-denuded zones near interfaces, reducing swelling threefold (0.4% vs. 1.2% in pure Nb). EELS confirms He concentrations of 63-96 He/nm³ in bubbles, corroborating the analytical calculation. In Zr/Nb multilayers, only 11% of implanted He contributes to bubbles, significantly lower than pure Nb (17.5%). This retention is asymmetrically distributed, with 7% in Zr layers (larger bubbles) and 4% in Nb layers (smaller bubbles and BDZs). The heterophase interfaces reduce He accumulation in Nb compared to its pure form, demonstrating their critical role in radiation tolerance. The classical MD and DFT simulations show that vacancies in Nb migrate toward interfaces (low 0.4 eV barriers), promoting recombination, while Zr traps defects with high barriers (~ 0.7 eV). He diffuses preferentially from the Nb side to the interface (0.19 eV barrier in Nb vs. 0.46 eV barrier in Zr). This asymmetry in the transport behavior of defects manifests as the observed BDZs. Zr/Nb multilayers exhibit 50% lower hardening ($\Delta H \approx 0.27$ GPa) than pure Zr/Nb due to interfacial defect annihilation. Sub-nm defects detected via VE-PALS also contribute to swelling/hardening. Tailoring layer thickness (<39 nm) and leveraging BCC/HCP interfaces optimize defect management. These insights advance the design of radiation-tolerant materials for nuclear reactors, with future work needed to validate them under high-temperature and neutron irradiation conditions. This work bridges microstructure-property relationships with atomic-scale mechanisms, offering predictive insights for advanced reactor components.

**Acknowledgments**
This work was financially supported by the European Union under the project Robotics and advanced industrial production (Reg. No. CZ.02.01.01/00/22_008/0004590). CzechNanoLab project LM2023051, funded by MEYS CR, is gratefully acknowledged for the financial support of the measurements/sample fabrication at LNSM Research Infrastructure. B. K. and H. S. S. acknowledge support from the UK Research and Innovation (UKRI)'s Engineering and Physical Sciences Research Council (EPSRC) through the Early Career Fellowship (EP/T026138/1). Computing facilities were provided by the Scientific Computing Research

Technology Platform (SC-RTP) at the University of Warwick. DFT calculations were performed using the Avon and Sulis HPC platforms, and Sulis is funded by the EPSRC Grant EP/T022108/1 and the HPC Midlands+ consortium. This work also used the ARCHER2 UK National Supercomputing Service (https://www.archer2.ac.uk) for performing the DFT calculations (through the Project e897). Parts of this research were carried out at ELBE at the Helmholtz-Zentrum Dresden – Rossendorf e. V., a member of the Helmholtz Association. We would like to thank the facility staff for their assistance. This work was partially supported by the Initiative and Networking Fund of the Helmholtz Association (FKZ VH-VI-442 Memriox) and the Helmholtz Energy Materials Characterization Platform (03ET7015).

# Asymmetrical Defect Sink Behaviour of HCP/BCC Zr/Nb Multilayer Interfaces: Bubble-Denuded Zones at Nb Layers

N. Daghbouj [a*], H.S. Sen [b*], M. BenSalem [a], J.Duchoň[c], B. Li [d], M. Karlík [e], F. Ge [f], V. Krsjak [g], P. Bábor[h], M.O. Liedke[i], M. Butterling[j], A. Wagner [i], B. Karasulu[b], T. Polcar[ak]

[a]Department of Control Engineering, Faculty of Electrical Engineering, Czech Technical University in Prague, Technická 2, 160 00 Prague 6, Czechia

[b]Department of Chemistry, University of Warwick, Coventry CV4 7AL, U.K.

[c]Institute of Physics of the Czech Academy of Sciences, Na Slovance 1999/2, 182 21 Prague 8, Czechia

[d]State Key Laboratory for Environment-friendly Energy Materials, Southwest University of Science and Technology, Mianyang, Sichuan 621010, China

[e]Department of Materials, Faculty of Nuclear Sciences and Physical Engineering, Czech Technical University in Prague, Trojanova 13, 120 00 Prague 2, Czechia

[f]Laboratory of Advanced Nano Materials and Devices, Ningbo Institute of Materials Technology and Engineering, Chinese Academy of Sciences, Ningbo 315201, China

[g]Institute of Nuclear and Physical Engineering, Faculty of Electrical Engineering and Information Technology, Slovak University of Technology, Ilkovicova 3, 812 19 Bratislava, Slovakia

[h]CEITEC - Central European Institute of Technology, Brno University of Technology, 616 00 Brno, Czech Republic

[i]Institute of Radiation Physics, Helmholtz-Zentrum Dresden-Rossendorf, Bautzner Landstr. 400, 01328 Dresden, Germany

[j]Reactor Institute Delft, Department of Radiation Science and Technology, Faculty of Applied Sciences, Delft University of Technology, Mekelweg 15, NL-2629 JB Delft, The Netherlands

[k]School of Engineering, University of Southampton, Southampton SO17 1BJ, United Kingdom

**Sample Structure**

Figure S1 presents a comprehensive microstructural and crystallographic characterization of the reference and multilayer samples used in this study. Panels (a–c) show transmission electron microscopy (TEM) images and high-resolution X-ray diffraction (HRXRD) data for the (110) single-crystal Nb reference. The distinct crystallographic orientation is confirmed by the sharp XRD peak at 38.5°, validating the single-crystalline nature of the Nb substrate. Panels (d) and (e) illustrate TEM micrographs of the polycrystalline Zr reference, revealing grains larger than 300 nm with multiple crystallographic orientations. The corresponding XRD spectra in panel (g) highlight a dominant (0002) texture near the free surface. Panels (h) and (i) provide automated crystal orientation mapping (ACOM) results of the Zr/Nb multilayer, indicating that grains are generally larger than individual layer thicknesses and exhibit strong textures of (0002) Zr and (110) Nb. These characterizations validate the microstructural stability and crystallographic orientation of the samples, which are critical for interpreting irradiation-induced effects.

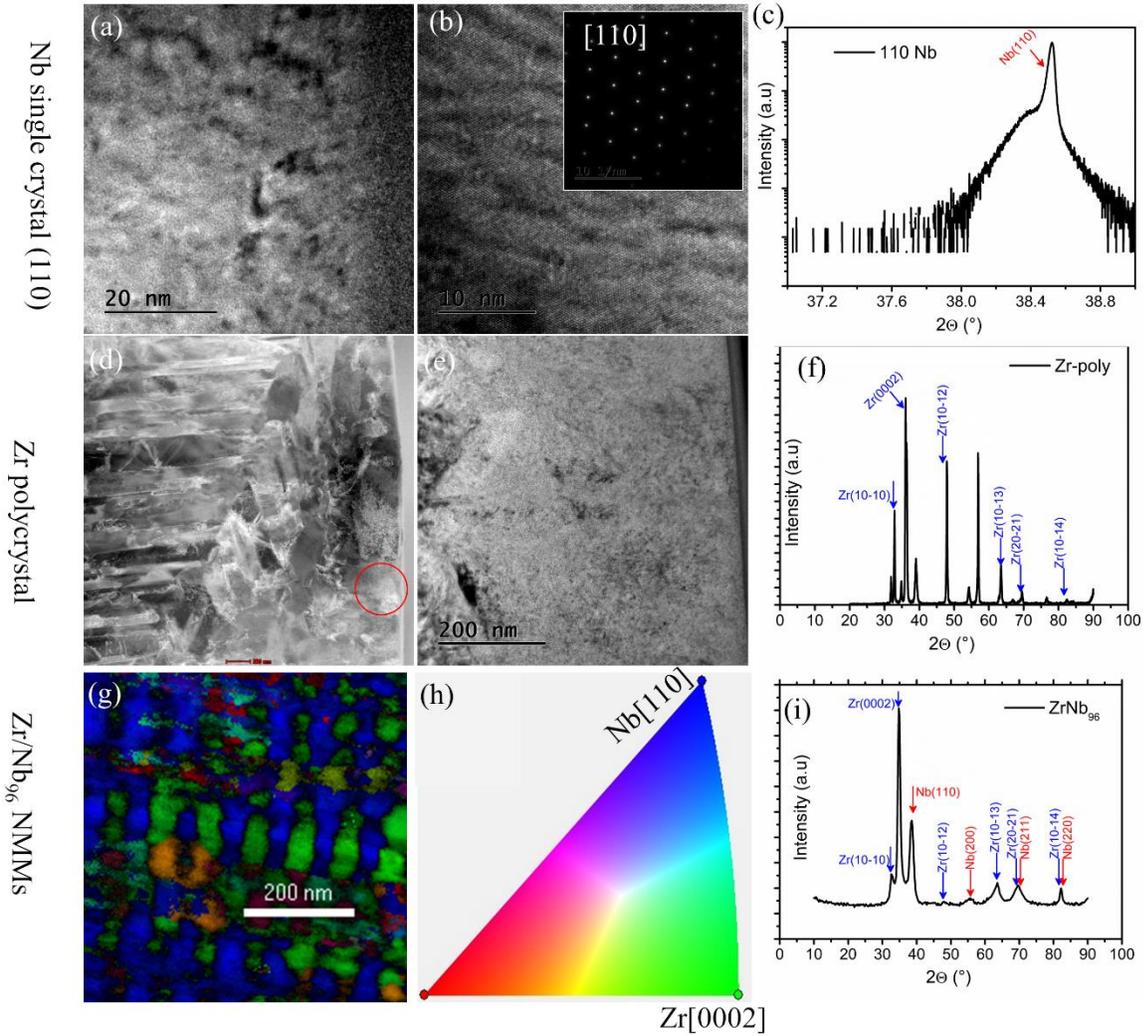

**Fig.S1.** Characterization of reference and multilayer samples. (a–c) TEM images and HRXRD pattern of the (110) single-crystal Nb reference, confirming distinct crystallographic orientation with the sharp XRD peak at 38.5°. (d, e) TEM micrographs of the polycrystalline Zr reference with an average grain size >300 nm, showing multiple grain orientations. (f, g) XRD spectra of polycrystalline Zr indicating a predominant (0002) orientation near the free surface. (h, i) Automated crystal orientation mapping (ACOM) of the Zr/Nb multilayer, revealing grains larger than the individual layer thicknesses and strong textures of (0002) Zr and (110) Nb.

**Ion irradiation**

The solid curves represent helium concentration profiles (in atomic%) predicted by SRIM and measured experimentally by SIMS (plotted against the right y-axis), while the dashed line corresponds to the simulated radiation damage profile, expressed in displacements per atom (dpa) and mapped on the left y-axis.

Key takeaways:

- For the Zr/Nb multilayers, SRIM predicts peak He concentrations of ~1 at.% (LF), 5 at.% (MF), and 12 at.% (HF), aligning with damage levels of ~0.3, 1.4, and 2.9 dpa, respectively, at a projected depth of 355 nm.
- The maxima of He concentration lie deeper than those of radiation damage.
- In elemental Zr and Nb, SRIM estimates He ions penetrate to approximately 380 nm and 250 nm, respectively.
- The SIMS profiles closely match the SRIM simulations, lending credibility to the modeled results.

This figure underscores the spatial decoupling between damage peaks and implanted He maxima and provides critical insight into the depth-dependent effects of ion irradiation in both pure and multilayer systems.

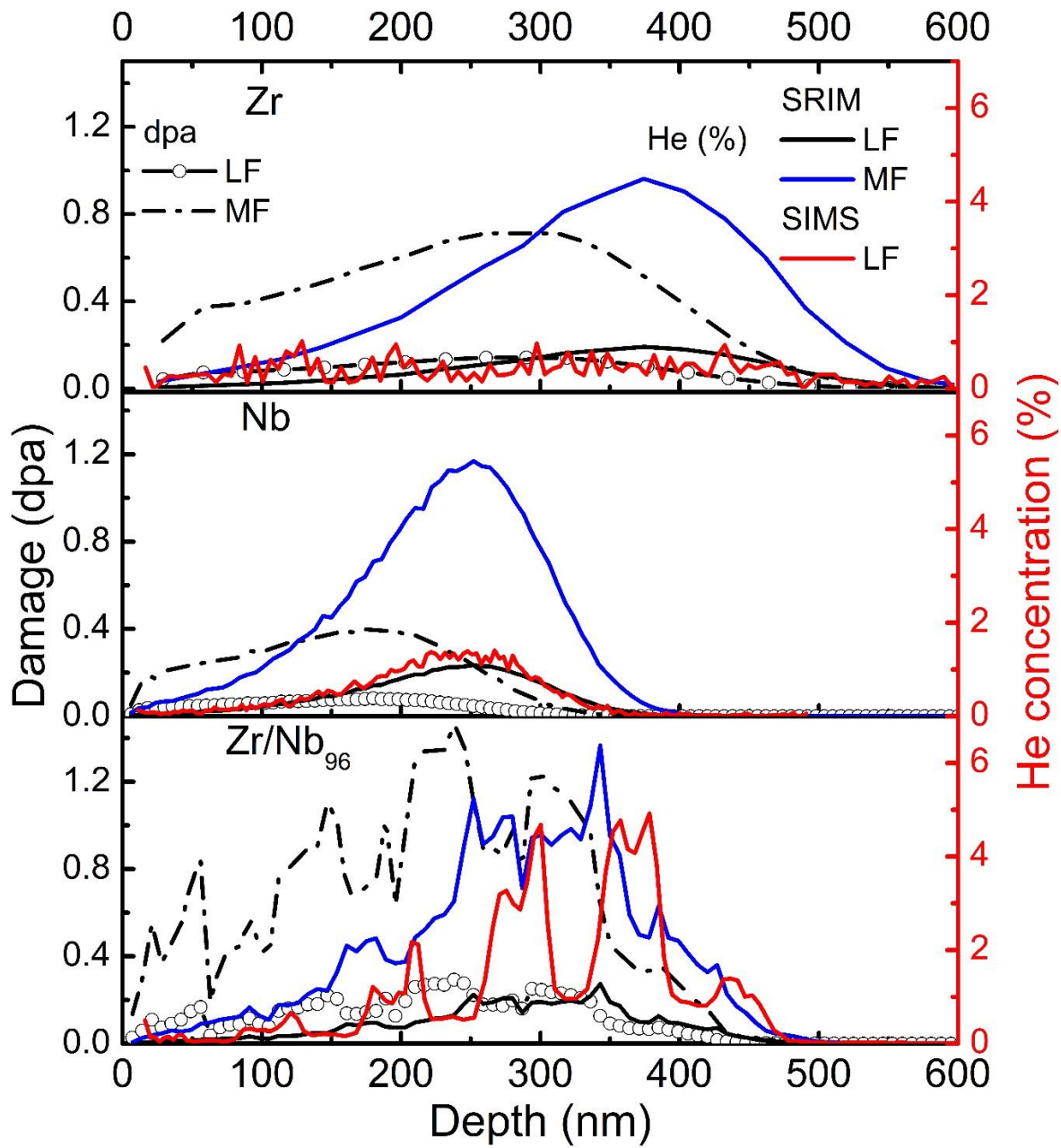

Fig. S2. Predicted (SRIM) and measured (SIMS) He concentration profiles (right y-axis) with corresponding radiation damage dpa (left y-axis) for (a) pure Zr, (b) pure Nb, and (c) Zr/Nb96 multilayers irradiated with 80 keV He$^+$ ions at low (LF = 1 × 10$^{16}$ He$^+$/cm²) and medium fluence (MF = 5 × 10$^{16}$ He$^+$/cm²).

**Statistic study**

We developed a series of algorithms in MATLAB® specifically designed for TEM images for the quantitative analysis of helium bubbles in irradiated materials. These algorithms enable the extraction of key morphological parameters, including bubble diameter, surface area, areal coverage, volume, and swelling. The workflow begins with image pre-processing, where raw TEM images are rotated and cropped to isolate the region of interest (ROI). To correct illumination heterogeneity, background subtraction is performed using singular value decomposition (SVD). This is followed by a sequence of image enhancement and segmentation steps, including adaptive thresholding, masking, morphological filtering, filling, and contour detection. Thresholding and masking parameters are tuned individually for each image to ensure optimal detection of all visible bubbles. Once segmentation is complete, statistical analysis is performed on the masked images to extract quantitative descriptors such as bubble count, size distribution, and swelling ratio. The extracted data are then averaged across each layer to generate depth-resolved profiles for different material phases (e.g., Zr and Nb), as illustrated in Fig. S2.

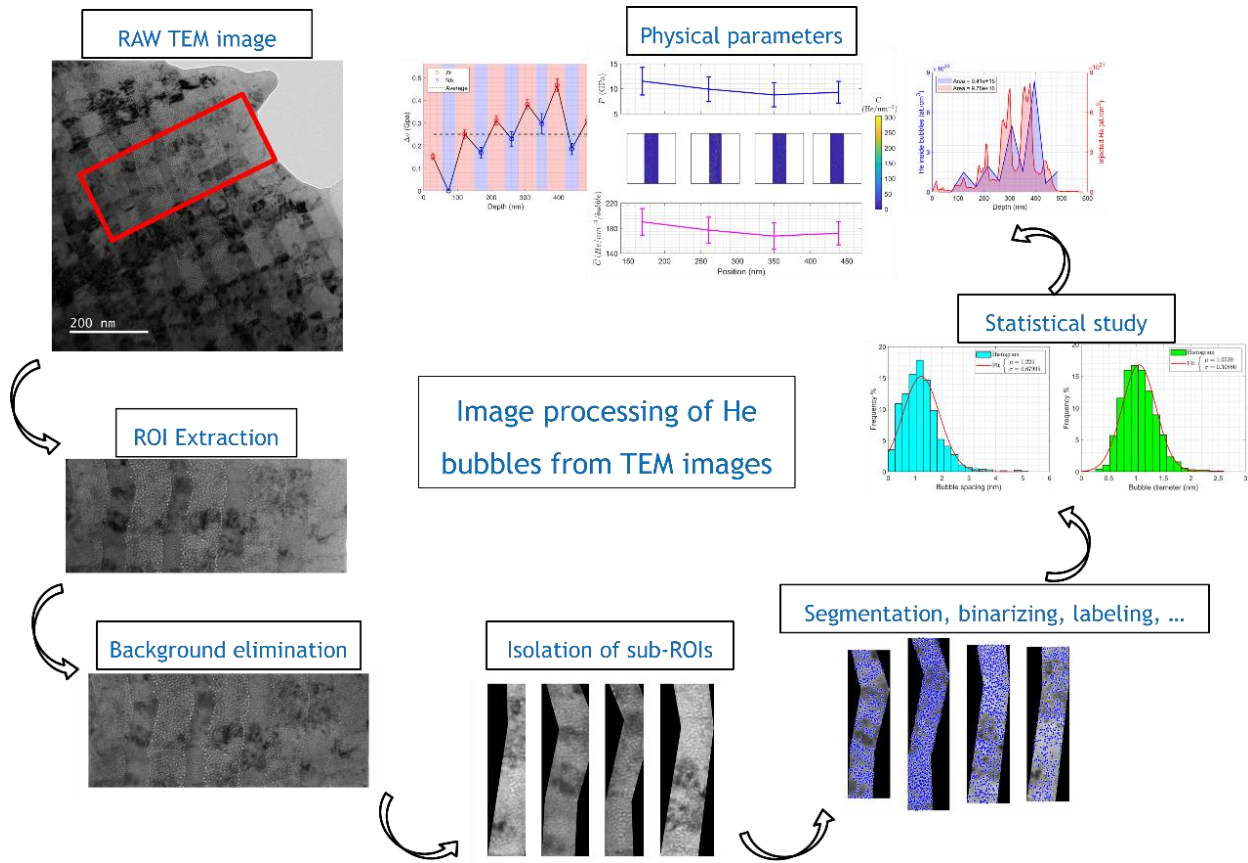

Fig. S3. Workflow of image processing and quantitative analysis of helium bubbles from TEM micrographs. The schematic illustrates steps of image pre-processing, segmentation, and parameter extraction, enabling determination of bubble diameter, density, swelling (S), bubble-induced hardening (σ), and helium-to-vacancy ratio (N).

**Nanoindentation**

Figure S2 details the nanoindentation measurement methodology and surface morphology characterization post-irradiation. Panels (a), (e), and (i) show continuous stiffness measurement (CSM) nanoindentation hardness vs. indentation depth curves for the Zr/Nb multilayers, single-crystal Nb, and polycrystalline Zr samples, respectively, with at least 16 indents performed per sample to ensure statistical confidence. Panels (b), (f), and (j) show corresponding hardness statistics, illustrating the reproducibility and consistency of measurements. Atomic force microscopy (AFM) images in panels (c), (j), and (k) demonstrate the negligible occurrence of pile-up or sink-in around indents within the 300–500 nm depth range. AFM surface roughness measurements in panels (d), (h), and (l) quantify post-irradiation surface roughness values between 6–8 nm for the multilayers and 9–14 nm for pure Nb and Zr crystals. These roughness levels do not significantly affect hardness measurements since indentation plastic zones exceed the surface roughness scale, ensuring reliable mechanical data despite irradiation-induced surface changes.

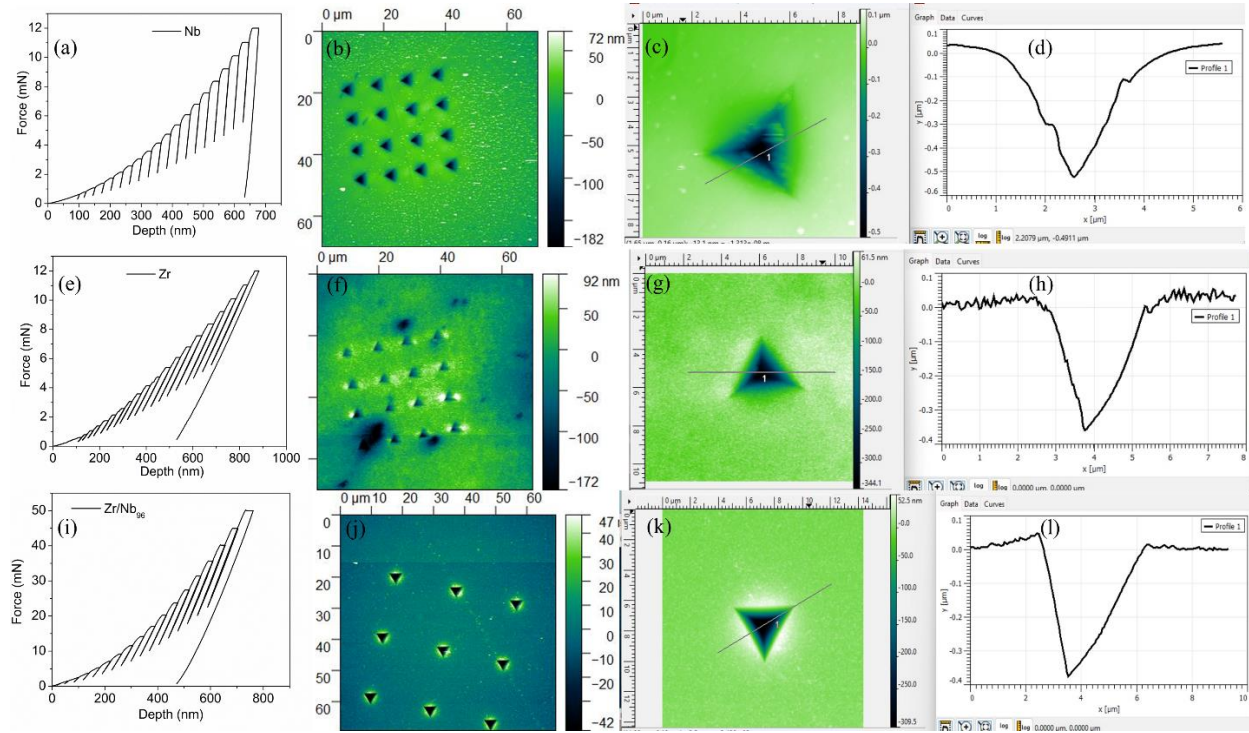

**Fig.S4.** Nanoindentation measurements and surface characterization. (a, e, i) Continuous stiffness measurement (CSM) hardness vs. indentation depth curves for Zr/Nb multilayers, single-crystal Nb, and polycrystalline Zr, respectively. (b, f, j) Hardness statistics showing minimal scatter across ≥16 indents per sample. (c, g, k) AFM images of indents confirming negligible pile-up or sink-in at 300–500 nm depth. (d, h, l) AFM surface roughness after irradiation, showing values of 6–8 nm for multilayers and 9–14 nm for pure Nb and Zr, which are insignificant relative to indentation depth.

Interface Model for Molecular Dynamics Simulations

Interface simulations were performed on a (110) Nb / (0002) Zr system, with a supercell measuring approximately 10 nm x 10 nm x 40 nm. The interface was constructed such that the [0002] direction of Zr and the [110] direction of Nb are oriented perpendicular to the interface plane (z-direction). To ensure coherent in-plane alignment, the [11$\bar{2}$0] (x-direction) and [1$\bar{1}$00] (y-direction) directions of Zr unit cell were aligned with the [001] and [1$\bar{1}$0] directions of Nb unit cell, respectively. The lattice vectors of both materials were then extended in the x and y directions to approximately 10 nm to enable the identification of matching (or nearly matching) lattice points at the interface. This allowed for the formation of a low-strain, semi-coherent interface between Nb and Zr. In the z-direction, each material slab was extended to a thickness of ~20 nm, resulting in a total supercell thickness of ~40 nm. A large simulation cell was used to minimize interactions between periodic images, particularly in regions containing structural damage or defects.

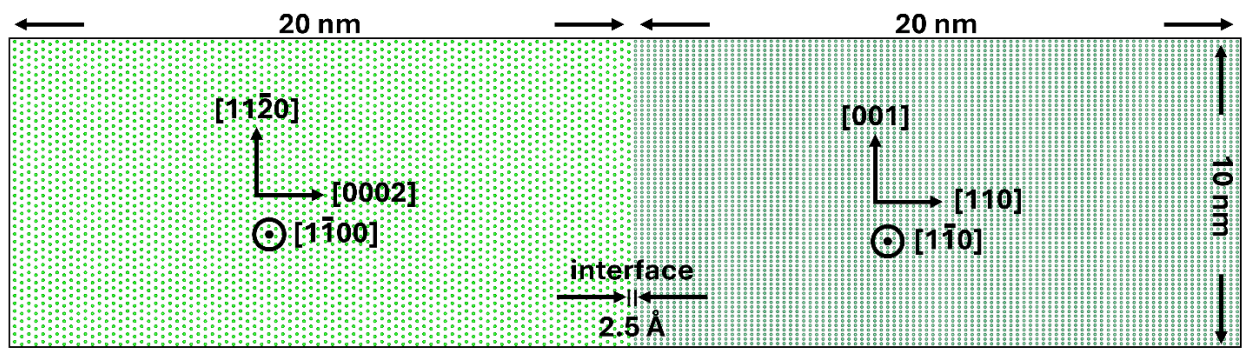

Fig S5: Interface model used for molecular dynamics simulations. The (110) Nb / (0002) Zr interface system was constructed with ~10 nm × 10 nm × 40 nm supercells to form a low-strain, semi-coherent interface. Each slab has a thickness of ~20 nm in the z-direction.

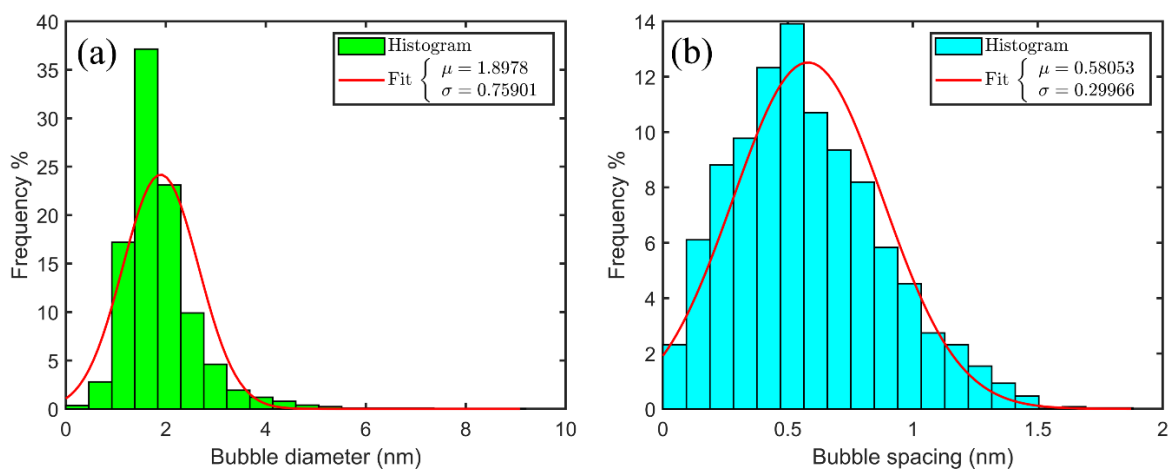

Fig. S6. Distribution within Zirconium of (a) bubble diameter and (b) bubble spacing with the Gaussian fit. The mean diameter is 1.9 nm, and the mean inter-bubbles distance is 0.6 nm.

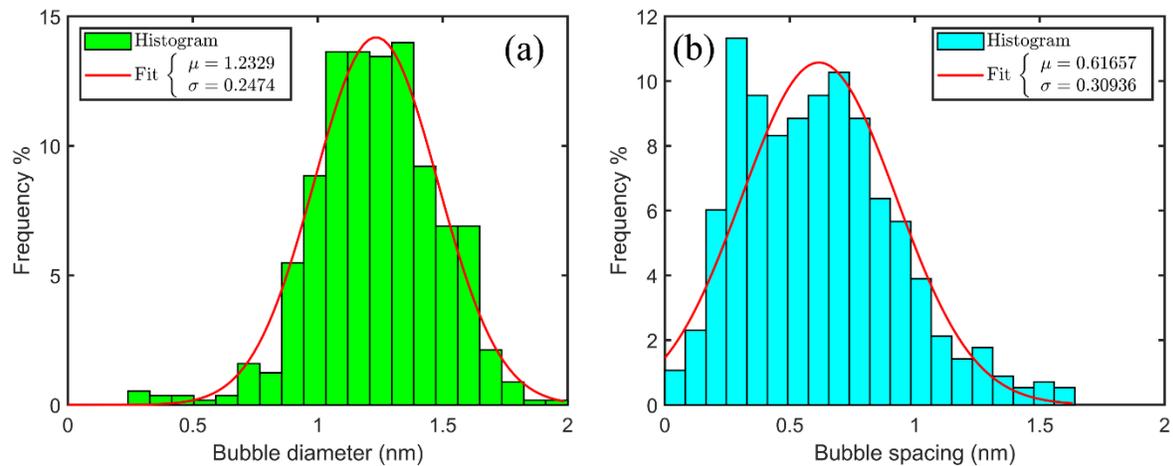

Fig. S7. Distribution within Niobium of (a) bubble diameter and (b) bubble spacing with the Gaussian fit. The mean diameter is 1.2 nm, and the mean inter-bubble distance is 0.6 nm.

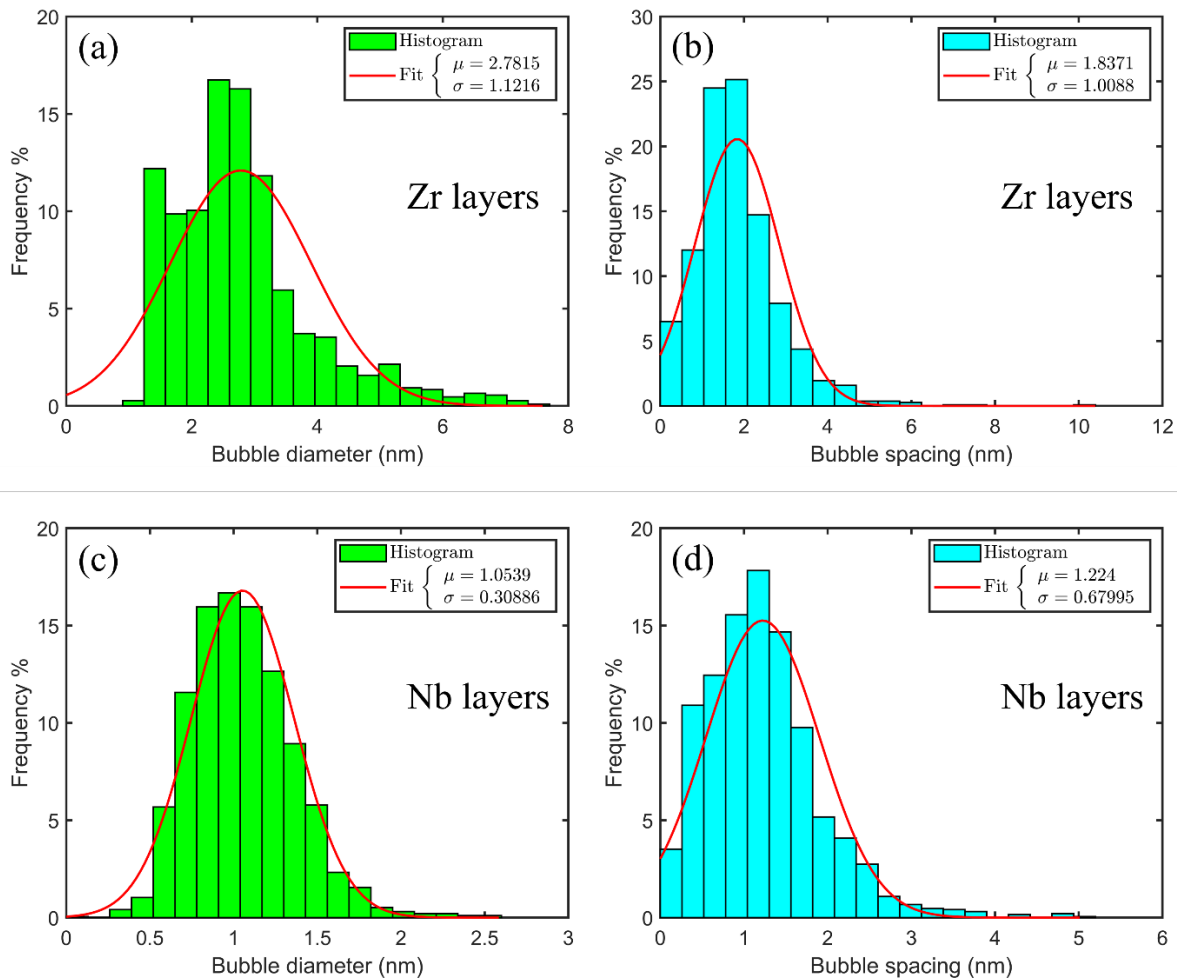

Fig. S8. Distribution within multilayer of (a) bubble diameter in Zirconium layers, (b) bubble spacing in Zirconium layers, (c) bubble diameter in Niobium layers, (d) bubble spacing in Niobium layers. The mean diameter of bubbles within Zirconium layers is 2.8 nm and 1 nm within Niobium layers. The mean bubble spacing within Zirconium layers is 1.8 nm and 1.2 nm within Niobium layers.

**Nix-Gao Model**

$$H = H_0\sqrt{\left(1 + \frac{h^*}{h_c}\right)} \Rightarrow H^2 = H_0^2\left(1 + \frac{h^*}{h_c}\right) \Rightarrow \underbrace{H^2}_{y} = \underbrace{H_0^2}_{a} + \underbrace{H_0^2 \times h^*}_{b}\underbrace{\frac{1}{h_c}}_{x}$$

$$\begin{cases} H_0 = \sqrt{a} \\ h^* = \dfrac{b}{a} \end{cases}$$

$H_0$: Bulk hardness in GPa.

$h^*$: Characteristic depth in nm.

$h_c$: Contact depth in nm.

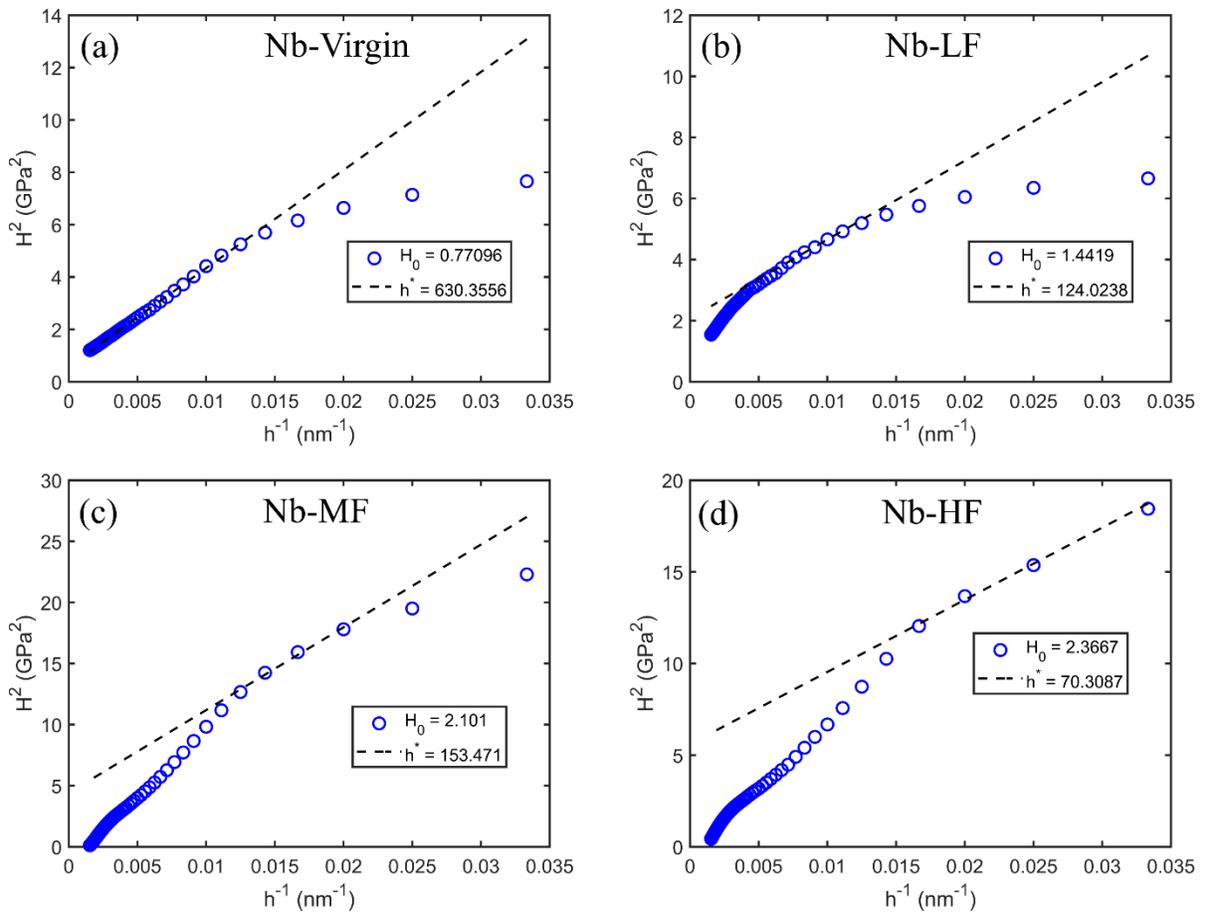

Fig. S9. Nix-Gao model fitting of Niobium squared Hardness function of inverse of contact depth: (a) virgin sample, (b) low fluence LF, (c) medium fluence MF, and (d) high fluence HF.

**Calculation of the Helium to vacancy ratio**

$$N_i = \frac{d_i \times N_A}{A_w}$$

$N_i$: Atomic density in $(nm^{-3})$.

$d$: Density in $(g/nm^3)$. For Zirconium $(6.52 \times 10^{-21} g/nm^3)$ and for Niobium is $(8.57 \times 10^{-21} g/nm^3)$.

$N_A$: Avogadro constant $(6.022 \times 10^{23})$.

$A_w$: Atomic weight in $(g/mol)$. For Zirconium $(91.224\ g/mol)$ and for Niobium is $(92.906\ g/mol)$.

$i$: Bubble index $i$.

$$N_{vacancy-i} = N_i \times V_{bubble}$$

$N_{vacancy}$: Number of vacancies.

$V_{bubble}$: Volume of bubble $i$, given by $(\frac{\pi}{6} D_i^3)$, with $D_i$ is the diameter of bubble $i$ in $(nm)$.

$$\overline{N}_{vacancy} = \frac{1}{m} \sum_{i=1}^{m} N_{vacancy-i}$$

$\overline{N}_{vacancy}$: Average number of vacancies, m number of bubbles for each region.

$$N_{He} = C_{He} \times V_{bubble}$$

$N_{He}$: Helium number per bubble.

$C_{He}$: Concentration of Helium per $nm^3$ per bubble.

$$He/V = \frac{N_{He}}{N_{vacancy-i}}$$

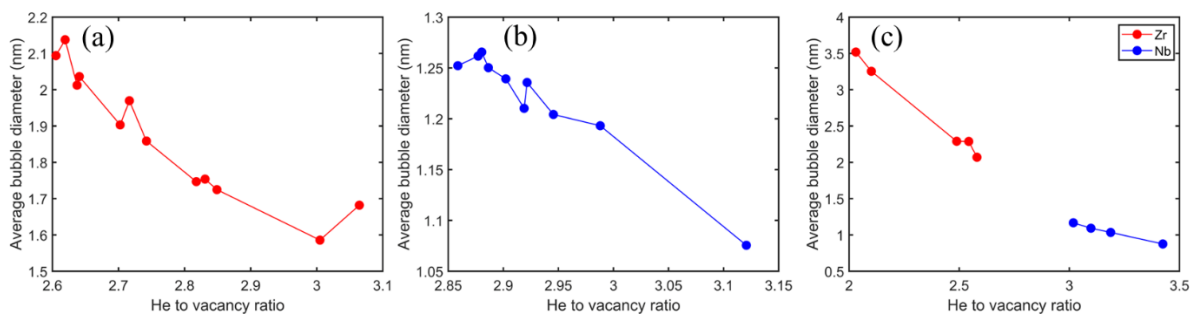

Fig. S10. Bubble diameter function of He to vacancy ratio (a) in Zirconium, (b) in Niobium, and (c) in Zirconium and Niobium inside all layers within Multilayer.

For comparison, studies on metals such as nickel-based alloys by Jäger et al. [1,2] and Walsh et al. [3] have shown that helium bubbles with a diameter of 2.0 nm in pure Ni, formed after helium implantation, can reach a maximum helium density of 2.2 He/vacancy, a value comparable to our results.

**Helium Detection by EELS**

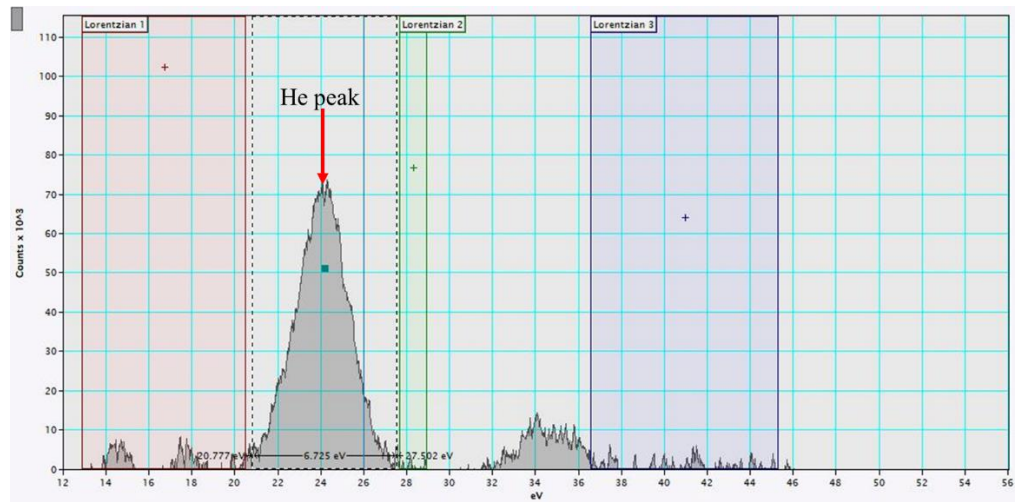

**Fig.S11.** Processed EELS spectra showing isolated helium K-edge peaks after background subtraction and plural scattering removal.

**Energy Barrier**

Table S1: Migration energy barriers towards the cluster of $V_4$ and the interface for a single vacancy ($V_1$), a He atom, and the $V_1He_1$ cluster.

| $E_m$ (eV) | $V_1$ | He | $V_1He_1$ |
|---|---|---|---|
| Nb bulk | 0.42 | 0.40 | 2.09 |
| Zr bulk – a | 0.68 | 0.61 | 1.46 |
| Zr bulk – c | 0.71 | 0.80 | 2.11 |
| Nb interface | 0.42 | 0.19 | 1.58 |
| Zr interface | 0.69 | 0.46 | 1.99 |